\documentclass[useAMS,usenatbib]{mn2e}
\usepackage{times} 
\usepackage{graphicx}
\usepackage{rotating}
\usepackage[mathscr]{euscript}
\usepackage{array}
\usepackage{epsfig}
\usepackage{amsmath}
\usepackage{natbib}
\def\mbh{\ifmmode{{\mathrm M}_{bh}\,}\else{M$_{bh}$\,}\fi}
\def\msigma{\ifmmode{{\mathrm M}_{bh}-{\sigma}\,}\else{M$_{bh}- \sigma$\,}\fi}
\def\msun{\ifmmode{{\mathrm M}_{\odot}}\else{M$_{\odot}$\,}\fi} 
\def\kms{\ifmmode{{\mathrm{km \, s^{-1}}}}\else{${\mathrm{km \, s^{-1}}}$}\fi}

\def\lesssim{\mathrel{\hbox{\rlap{\hbox{\lower4pt\hbox{$\sim$}}}\hbox{$<$}}}}
\def\gtrsim{\mathrel{\hbox{\rlap{\hbox{\lower4pt\hbox{$\sim$}}}\hbox{$>$}}}}

\begin{document}
\title[Orbital spectral analysis of the stellar halo]{Probing the shape and history of the Milky Way halo with orbital spectral analysis}

\bigskip

\author[Valluri, M. et al. ]{ Monica Valluri $^{1}$\footnote{E-mail:mvalluri@umich.edu; vpdebattista@uclan.ac.uk}, Victor P. Debattista $^{2,3}$, Thomas R. Quinn $^{4}$, Rok Ro\v{s}kar $^{5}$ and James Wadsley $^{6}$\\
$^{1}$ Department of Astronomy, University of Michigan, Ann Arbor, MI 48109, USA\\
$^{2}$ Jeremiah Horrocks Institute, University of Central Lancashire,  Preston, PR1 2HE, UK\\
$^{3}$ RCUK Fellow\\
$^{4}$ Astronomy Department, University of Washington, Box 351580, Seattle, WA 98195-1580, USA\\
$^{5}$ Department of Theoretical Physics, University of Z\"urich, Winterthurerstrasse 190, CH-8057 Z\"urich, Switzerland\\
$^{6}$Department of Physics and Astronomy, McMaster University, Hamilton Ontario, L8S 4M1, Canada}
\paperheight = 11.in
\textheight = 8.0in

\date{\today}
\pagerange{\pageref{firstpage}--\pageref{lastpage}} \pubyear{2011}

\maketitle

\label{firstpage}

\begin{abstract} 
Accurate phase space coordinates (three components of position and
velocity) of individual halo stars are rapidly becoming available with
current and future surveys such as SDSS-SEGUE, RAVE, LAMOST,
SkyMapper, HERMES and eventually {\it Gaia}.  This will enable the
computation of the full 3-dimensional orbits of these stars.  Spectral
analysis of halo orbits can be used to construct ``frequency maps''
which provide a compact representation of the 6-dimensional phase
space distribution function. A frequency map readily reveals the most
important major orbit families in the halo, and the orbital abundances
in turn, reflect the shape and orientation of the dark matter halo
relative to the disk.  We apply spectral analysis to halo orbits in a
series of controlled simulations of disk galaxies. Although the shape
of the simulated halo varies with radius, frequency maps of local
samples of halo orbits confined to the inner halo contain most of the
information about the global shape of the halo and its major orbit
families. Quiescent or adiabatic disk formation results in significant
trapping of halo orbits in resonant orbit families (i.e. orbits with
commensurable frequencies). If a good estimate of the Galactic
potential in the inner halo (within $\sim 50$~kpc) is available, the
appearance of strong, stable resonances in frequency maps of halo
orbits will allow us to determine the degree of resonant trapping
induced by the disk potential. The locations and strengths of these
resonant families are determined both by the global shape of the halo
and its distribution function. Identification of such resonances in
the Milky Way's stellar halo would therefore provide evidence of an
extended period of adiabatic disk growth.  If the Galactic potential
is not known exactly, a measure of the diffusion rate of large sample
of $\sim 10^4$ halo orbits can help distinguish between the true
potential and an incorrect potential. The orbital spectral analysis
methods described in this paper provide a strong complementarity to
existing methods for constraining the potential of the Milky Way halo
and its stellar distribution function.
\end{abstract}



\section{Introduction}
\label{sec:intro}

Over the last decade a coherent picture of the formation of the
stellar halo of the Milky Way (hereafter MW) has begun to emerge both
from observational surveys of the Galaxy and cosmologically motivated
simulations. In this ``concordance cosmological model'' galaxies are
embedded in dark matter halos and form via the merger of protogalactic
fragments. The fragments consist of dark matter, gas, and the first
generations of stars which formed in high density peaks in the early
Universe. An extended period of ``late infall'' including the
accretion and tidal shredding of numerous dwarf satellite galaxies
\citep{bullock_johnston_05, delucia_helmi_08, font_etal_08,
  johnston_etal_08, cooper_etal_10} has continued to build up the halo
even at the present time. In contributing to the growth of disk
galaxies, gas accretion competes with accretions and mergers of
satellites which tend to thicken and disrupt the disk.  Recent
resolved-star all-sky surveys such as SDSS-SEGUE~\citep{SEGUE}, and
RAVE~\citep{steinmetz_etal_06} have revealed that current observations
of the stellar halo are broadly consistent with the $\Lambda$CDM
galaxy formation paradigm.  For example the detection in SDSS data of
numerous substructures in the MW halo in the form of tidal streams
\citep{newberg_etal_02, yanny_etal_03, belokurov_etal_06field}, the
measurement of the degree of clumpiness in the distribution of stars
in the stellar halo \citep{bell_etal_08}, and the discovery of
numerous ultra-faint dwarf spheroidal galaxies
\citep{willman_etal_05ursa, zucker_etal_06canes, belokurov_etal_07},
have all bolstered evidence that the Milky Way's stellar halo was
produced in the manner consistent with $\Lambda$CDM.

Although this picture of galaxy formation has been very successful,
the availability of phase space coordinates for tens of thousands of
halo stars from resolved star surveys have revealed the inadequacies,
and inherent degeneracies, of current analysis tools. For instance
phase space data are currently compared with simulations using simple
measures such as orbital eccentricity \citep{sales_etal_09}, not only
yield degenerate results \citep{wilson_etal_10,dierickx_etal_10}, but
depend on assumptions regarding the potential of the Galaxy.  Although
gas accretion probably dominates in disk galaxies it is unclear from
the current data whether the MW's inner stellar halo, and thick disk
formed purely {\it in situ } \citep{schoenrich_binney_09,
  loebman_etal_11} or whether a pre-existing thin disk was heated
\citep{villalobos_helmi_08,kazantzidis_etal_08} or if they were
entirely created by satellite accretion \citep{abadi_etal_03,
  yoachim_dalcanton_08}. There is also evidence that the inner stellar
halo may be smoother (less {\it rms} density variation) and more metal
rich than the outer halo \citep{helmi_etal_11}. In addition, a recent
analysis of the full space motions of $\sim 17,000$ halo stars from
the SDSS-SEGUE calibration sample \citep{carollo_etal_07,
  carollo_etal_10, beers_etal_11} suggests that the Galactic halo
consists of two overlapping components: an inner halo, which is
rotating in the same direction as the disk and an outer halo, with a
small retrograde motion. In addition the two components have different
density profiles, stellar orbits and metallicites \citep[however
  see][for a different view.]{schoenrich_etal_11}.

Current analysis tools rely on comparing the average distributions of
stars in the halo with those arising from simulations are not powerful
enough to compare the full 6-dimensional phase space distribution
functions (DF) of $10^5-10^6$ halo stars and the information
on their metallicities and ages that will become available following
the launch of {\it Gaia} \citep{perryman_etal_01}. It is impossible
for current analyses to rule out the possibility that a significant fraction
of nearby halo stars are not formed in our own potential, and have
been heated up into a very thick spheroidal inner halo.  Addressing
questions of this kind are of key importance when trying to constrain
galaxy formation models, and in uncovering the formation history of
the Milky Way, and can only be achieved with new techniques that can
quantitatively compare the self-consistent chemo-dynamical and
phase-space distribution function  of the stellar halo
with those from models.

In this paper we show that frequency analysis tools can be applied to
orbits of halo stars to uncover the phase space distribution function
of the entire stellar halo. These methods can be applied to both the
real DF of MW halo orbits and to those from $N$-body simulations,
permitting detailed quantitative comparisons of the orbit families in
the distribution functions. We also show that the halo orbit DF
(represented by a frequency map) reflects the global shape of the dark
matter halo and its orientation relative to the disk.

One of the robust predictions of the $\Lambda$CDM cosmological
paradigm is that the dark matter halos of galaxies like the MW are
triaxial \citep{dubinski_carlberg_91, jing_suto_02,
bailin_steinmetz_05, allgood_etal_06}. Orbits of particles in triaxial
potentials are very different from those in spherical and axisymmetric
potentials. For instance none of the orbits conserve any component of
angular momentum. The fraction of such ``triaxial orbits'' in a
potential is a strong indicator of the global shape of the
potential. If halos are elongated in ways that reflect their accretion
history and their orientation relative to large scale filaments
\citep{helmi_etal_11}, determining the shape and the orbital
populations could give us important clues to the formation of the MW.

Recent simulations have shown that when baryons cool and condense at
the centers of triaxial dark matter halos, the halos become more
axisymmetric especially at small radii \citep{dubinski_carlberg_91,
  kazantzidis_etal_04shapes, deb_etal_08, tissera_etal_10,
  kazantzidis_etal_10}. Two recent studies \citep{deb_etal_08,
  valluri_etal_10} analyzed the orbital properties of halo particles
in a series of $N$-body simulations in which baryonic components were
grown adiabatically in triaxial and prolate halos.  \citet[][hereafter
  V10]{valluri_etal_10} showed that although the inner one-third of
the halo becomes nearly oblate following the growth of a baryonic
component, only a  fraction of the orbits change their true
orbital characteristics. When the baryonic component is a disk galaxy,
most of the halo orbits retain a memory of their orbital
actions. Since the actions are adiabatic invariants, even nearly
oblate halos can have a significant fraction of box orbits and
long-axis tubes.

This ability of orbits to retain a memory of their initial conditions
has been previously noted in several numerical studies: when the
potential is changing fairly violently e.g. during a major merger
\citep{valluri_etal_07} or during the tidal disruption of a satellite
in the field of a larger galaxy \citep{helmi_dezeeuw_00}, stellar
orbits largely conserve their integrals of motion. In the latter case
\citet{helmi_dezeeuw_00} showed that the orbital actions can be used
to recover the relics of the tidal disruption of a dwarf satellite by
the MW potential.  Recently a number of studies
\citep{mcmillan_binney_08, gomez_helmi_10, gomez_etal_10} have shown
that the orbital frequencies $\Omega_R$ (radial oscillation or
epicyclic frequency) and $\Omega_\phi$ (tangential oscillation or
rotation frequency) of stars belonging to a single satellite are
tightly correlated and that this correlation remains strong long after
the remnant has become completely well mixed in configuration
(physical) space. Consequently, correlations between orbital
frequencies can be used to identify stars that belong to individual
accretion events, and simultaneously determine the true galactic
potential and the time since each satellite galaxy was disrupted
\citep{mcmillan_binney_08, gomez_helmi_10}.

Following the method outlined by \citet{carpintero_aguilar_98}, V10
\citep[also][]{deibel_etal_11} used the oscillation frequencies of
orbits of particles in simulated halos to classify orbits into major
families. V10 also used orbital frequencies to quantify the shapes of
orbits (and to relate the orbital shapes to the shapes of the halos),
to identify orbits which are chaotic, and to identify important
resonances (regions of phase space occupied by orbits with
commensurable orbital frequencies). In this paper we show that the
{\it relative distribution of orbital types and the identification of
  important resonances} populated by halo orbits strongly reflects the
orientation of the triaxial halo relative to the galactic disk. We
will show that the discovery of large numbers of halo stellar orbits
trapped in resonances could put constraints on the form of the
Galactic potential and the DF of the stellar halo.

Why are resonances important? When a time-dependent force acts on a
system like a galaxy, long-lived resonant interactions play an
important role in the evolution of the system and leave imprints in
its phase space structure. The identification of stars trapped in
resonances can put constraints on the potential and its components
(e.g. the bar and spiral arms). Resonant interactions between
individual orbits and a changing potential influence time-dependent
(secular) evolution: e.g. it can cause stars to ``levitate'' to form a
thick disk \citep{sridhar_touma_96}, it may result in the formation of
polar rings or counter-rotating disks \citep{tremaine_yu_00}, and it
can cause resonant shocking (or torquing) of stars in satellites as
they are disrupted in dark matter halos
\citep{choi_etal_09}. ``Capture into resonance'' is studied
extensively in the context of planetary systems, where a migrating
planet can capture planets or planetesimals into mean motion
resonances \citep{malhotra_93,yu_tremaine_01}.  In the planetary
dynamics literature it has been shown that resonant trapping of
planetesimals occurs in slowly varying potentials, but can be
prevented when the drift or migration rate is sufficiently high
(i.e. non adiabatic) \citep[e.g.][]{quillen_06}. The identification of
a significant fraction of resonantly trapped halo orbits could
therefore provide clues to the way in which the halo potential has
changed over time.

Laskar's frequency analysis method is particularly good at identifying
resonances \citep{robutel_laskar_01}. We show that when the method is
applied to orbits in a self-consistent distribution function, the
method permits easy identification of the major orbit families and an
assessment of the relative importance of each family to the phase
space DF.

This paper is organized as follows: Section~\ref{sec:method} describes
the simulations analyzed in this paper and briefly describes the
frequency analysis of orbits. Section~\ref{sec:fmaps} presents the
results of our analysis of adiabatic simulations of isolated galactic
potentials consisting of both axisymmetric and triaxial dark matter
halos with particle disks of various orientations \citep[taken mostly
  from][hereafter D08]{deb_etal_08}. We also discuss the effects of
disk galaxies which form self-consistently from hot halo gas in a
spherical or prolate halo. These simulations include the
hydrodynamical effects of gas cooling, star formation and supernova
feedback \citep{stinson_etal_06, roskar_etal_08}. In
Section~\ref{sec:potential} we show how the mean orbital diffusion
rates of a large ensemble of orbits selected from a distribution
function can be used to assess how much it deviates from
self-consistent equilibrium and discuss how this measure is affected
when the assumed Galactic potential is incorrect. In
Section~\ref{sec:discuss} we summarize our results, and discuss their implications
for future large data sets which will obtain the 6 dimensional
orbits of stars in the MW halo.


\section{Simulations and Numerical Methods}
\label{sec:method}

\subsection{Simulations}
\label{sec:simulations}

\begin{table*}
\begin{centering}
\begin{tabular}{ccccccll}\hline 
\multicolumn{1}{c}{Run } &
\multicolumn{1}{c}{$r_{200}$} &
\multicolumn{1}{c}{$M_{200}$} &
\multicolumn{1}{c}{$M_b$} &
\multicolumn{1}{c}{$f_b$}&
\multicolumn{1}{c}{$t_g$} &
\multicolumn{1}{c}{Run Description }&
\multicolumn{1}{c}{Reference} \\ 
 Name &  [kpc] & [$10^{12} \msun$] & [$10^{11} \msun$] &       & [Gyr] &    &        \\ \hline
SNFWD   &    85       &      0.66                         &              0.66                &  0.1        &     5    & {\em Spherical halo+stellar disk}&  This paper    \\
SA1    &   215  &   4.5 & 1.75  & 0.039   &  5  &{\em Triaxial halo+short-axis stellar disk} & D08 \\ 
LA1     & 215   &   4.5 & 1.75  & 0.039  &  5   &{\em Triaxial halo+long-axis stellar disk}   &  D08 \\
IA1      & 215   &   4.5 & 1.75  & 0.039   &  5   &{\em Triaxial halo+intermediate (y)-axis stellar disk}   & D08\\
TA1     & 215   &   4.5 & 1.75  & 0.039  &  5   & {\em Triaxial halo+tilted stellar disk}    & D08\\
SA2    & 215   &   4.5 &  0.52 & 0.012   &  1.5   & {\em Triaxial halo+barred stellar disk} &  D08 (there labelled BA1) \\
\hline
\end{tabular}
\end{centering}
\caption{The collisionless simulations in this paper.  $M_b$ is the
  mass in baryonic disk and $f_b$ is the baryonic mass fraction.
  $t_g$ is the time during which the baryonic disk is grown.  In all
  the models the exponential disk has a radial scale length of
  3~kpc. The last column contains the references where the simulation
  was first reported.}
\label{tab:simulations}
\end{table*}

We analyzed two types of controlled simulations (a) $N$-body
simulations in which an exponential, thin stellar disk (consisting of
collisionless particles) was grown adiabatically inside an isolated
halo; (b) $N$-body+SPH hydrodynamical simulations of the formation of
stellar disks from initially hot gas distributed inside a spherical or
prolate halo and allowed to cool and form a disk of gas in which stars
form. In general we assume that the potential of the galaxy model is
perfectly known, and we characterize the distribution function, its
orbital properties, and its dependence on the radial variation of the
shape of the halo and its orientation relative to that of the disk. In
practice, the potential and the DF of the halo need to be determined
simultaneously, or (in the absence of adequate kinematical constraints)
 the potential alone will be determined, within some
radius. We therefore ran a few additional simulations, designed to
determine if it is possible to constrain the potential from the
orbital properties of an ensemble of halo orbits.

In the controlled simulations presented in this paper, initially
spherical isotropic NFW \citep{nfw} halos were generated via
Eddington's formula \citep[][\S~4.3.1]{BT} with each halo composed of
two mass species arranged on shells. The inner shell has less massive
particles than the outer one, which allows for higher mass resolution
at small radii. Most of the dark matter particles in the inner part of
the halo have masses of $10^6$\msun.  Prolate and triaxial halos
(consisting of $4\times 10^6$ particles) were generated via mergers of
the spherical NFW halos (see D08 for details).

In the simulations, disks of particles were grown (starting from
nearly zero initial mass) adiabatically and linearly on a timescale
$t_g$ inside a dark matter halo (see Tab.~\ref{tab:simulations} for
details of parameters of the simulations). Disks were grown in a
spherical halo (model SNFWD) and in triaxial halos with the disk
plane oriented in various ways relative to the principle axes
of the halo: (a) perpendicular to the short axis (model SA1), (b)
perpendicular to the long axis (model LA1), (c) perpendicular to the
intermediate axis (model IA1), and (d) tilted at an angle of
30$^\circ$ to the $x-y$ plane of the triaxial halo, by rotating it
about the $y$ axis (model TA1)\footnote{Note that $x, y, z$ are
defined to be the long, intermediate and short axes respectively, of
the triaxial halo.}.  D08 showed that in all of these models, the
shape of the halo within the inner 1/3 of the virial radius becomes
nearly (but not exactly) oblate following the growth of the disk, with
the short-axis of the oblate part of the halo co-aligned with the spin
axis of the disk. In most of the simulations the disk particles remain
stationary, hence the disk is rigid throughout. In one case (model
SA2), a ``light'' disk of particles was made ``live'' after the disk
had grown to its final mass.  This disk subsequently formed a bar which
persisted for $\sim 10$~Gyr before dissolving because of the triaxial
halo \citep{berentzen_shlosman_06}.  Additional details may be found
in D08. All the collisionless simulations were evolved with {\sc
pkdgrav}, an efficient, multi-stepping, parallel tree code
\citep{stadel_phd}. Dark matter particles had a softening parameter
$\epsilon = 0.1$~kpc, and that of stars was in the range $\epsilon =
60-100$~pc.

Two of our controlled simulations contain a baryonic (gas+star) disk,
which forms self-consistently from hot gas in a spherical (model
SNFWgs) or prolate (model SBgs) halo.  The baryonic component is 10\%
of the total mass and initially has the same density distribution as
the dark matter particles. The halo and gas particles are given an
initial specific angular momentum $j$, determine by overall
cosmological spin parameter $\lambda=(j/G)(|E|/M^3)^{1/2}= 0.039$,
which is motivated by cosmological $N$-body experiments
\citep{bullock_etal_01}.  Both the spherical halo and the progenitor
halos that were merged to produce a prolate halo had the same
angular momentum parameter $\lambda$. Each component is modeled with
$10^6$ particles, with the dark matter particles of mass $10^6$~\msun,
and gas particles having an initial mass of $10^5$~\msun. The gas
particles are allowed to cool and form stars of typical mass around
$3\times 10^4$~\msun following the prescription in
\citet{stinson_etal_06}. The net angular momentum allows the gas to
form a disk as it cools, resulting in a stellar disk as star formation
occurs. The simulation closely follows that described in
\citet{roskar_etal_08} and is evolved with the parallel $N$-body+SPH
code GASOLINE \citep{gasoline} for 10 Gyr.

Although controlled simulations are useful for testing the effects of
disks with different orientations on the halo DF, these models are not
fully realistic depictions of how disk galaxies form. In the current
hierarchical structure formation paradigm, disk galaxies probably
experienced several gas rich mergers, at least in their early history,
and continue to accrete small satellites today which add to the
stellar halo. We defer the study of stellar halos from cosmological
simulations drawn from the MUGS project \citep{stinson_etal_10} to a
future paper \citep{valluri_etal_11_mugs}.

\subsection{Selecting the halo orbit samples}
\label{sec:orbits}

In each of the simulations we selected $1-2\times 10^4$ dark matter
particles. The particles were either randomly distributed within some
spherical volume of radius $r_g$ centered on the model's galactic
center, or within a region of radius $R_s$ from the location of the
``sun'' (which was assumed to be at 8~kpc from the Galactic
center). Since most of the potentials we studied are non-axisymmetric,
the azimuthal location of the ``sun'' in the equatorial plane relative
to the major axis of the triaxial halo is an additional
parameter. Rather than choosing a specific (arbitrary) angle in
azimuth for the solar location, we select particles within a ``torus''
of width $R_s$ and with radius 8~kpc. When we selected subsamples of
the torus region we found that the results did not depend much on the
precise azimuthal location of the ``sun'', so long as the radial
region sampled was 10~kpc in size, or larger.

Following V10 we studied orbits of halo particles after the disk had
grown to its final mass and the halo had relaxed to a new equilibrium.
The orbits were integrated for $50$~Gyr in a frozen potential
corresponding to the new equilibrium potential given by the full mass
distribution of the simulation (dark matter and baryons) using an
integration scheme based on the {\sc PKDGRAV} tree. The $50$~Gyr
integrations are used only because they yield highly accurate
frequencies for the halo orbits with the longest periods, and should
not be construed to imply that this is a physically meaningful time
period. (In a similar way, it is possible to integrate orbits of MW
halo stars in a fixed potential for equally long times in order
to characterize the nature of their {\it current} orbit.) None of the
controlled simulations include a stellar halo, so we assume that the
dark matter halo orbits can be used to represent the orbits of
particles in the stellar halo. While this is clearly not an ideal
assumption, many studies of the stellar halo use prescriptions to
``tag'' the most tightly bound dark matter particles in dwarf
satellites as ``star particles'' \citep[e.g.][]{bullock_johnston_05,
  cooper_etal_10}. Elsewhere we analyze a fully cosmological
hydrodynamical simulation of a disk galaxy from the MUGS project
\citep{stinson_etal_10} and show that in the inner halo, (a region on
which we focus) star particles and dark matter particles have similar
orbital properties, justifying this assumption
\citep{valluri_etal_11_mugs}.

\subsection{Laskar Frequency Mapping}
\label{sec:laskar}

In three dimensional Hamiltonian potentials, the phase space structure
of regular orbits can be described by three actions $J_\alpha$
($\alpha = 1,2,3$) and three angle variables $\theta_\alpha$, which
constitute a canonically conjugate coordinate system. The actions are
integrals of motion and are conserved along the orbit, while the
angles increase linearly with time. The angle variables at any time
$t$ are given by $\theta_\alpha(t) = \theta_\alpha(0)+\Omega_\alpha
t$, where the $\Omega_\alpha$ are called ``fundamental frequencies''.
The actions are adiabatic invariants and consequently remain constant
as the potential of the system changes adiabatically.  Regular orbits
in 3-dimensional potentials can therefore be thought of as occupying
the surfaces of 3 dimensional tori, with the size of each torus
characterized by the actions $J_\alpha$, while the angle variables
$\theta_\alpha(t)$ represent the traversal of the orbit over the
surface of the torus in each dimension.

Since the angle-action coordinates are related to the classical
spatial coordinates and momenta via a coordinate transformation, it
can be shown that traditional space and velocity coordinates can be
represented by a time series of the form: $x(t) = \sum A_{k}
e^{i\omega_kt}$, and similarly for other phase space coordinates
e.g. $y(t), v_x(t)$ (where the sum is over all terms in the
spectrum). A Fourier transform of such a time series will yield the
spectrum of orbital frequencies $\omega_k$ and associated amplitudes
$A_k$, that govern the motion of the orbit \citep{binney_spergel_82,
  binney_spergel_84,BT}. For most regular orbits only three of the
frequencies in the spectrum (of a given orbit) are linearly
independent, (i.e. all other frequencies $\omega_k$ can be expressed
as $\omega_k = l_k\Omega_1+m_k\Omega_2+n_k\Omega_3$, where $l_k, m_k,
n_k$ are integers).  $\Omega_1, \Omega_2, \Omega_3$ are nothing other
than the ``fundamental frequencies'' described above and the
associated 3 orbital actions $J_1, J_2, J_3$ can be computed from the
amplitudes $A_k$.

If all orbits in the potential are regular and the DF can be written
as a continuous function of global actions $J_1, J_2,J_3$ e.g.
$f(J_1,J_2,J_3)$ then the corresponding fundamental frequencies and
the associated angle variables $\theta_1, \theta_2, \theta_3$ vary in
a continuous manner across phase space. However it is only possible to
write the DF as a function of global actions for special cases
e.g. strictly axisymmetric potentials or separable triaxial
(St\"ackel) potentials. If analytic global actions are known then one
can simply use surfaces of section to map the phase space structure
and to identify resonant orbits and chaotic orbits (BT08).  In the
study of three dimensional orbits in realistic potentials analytic
integrals of motion other than the Hamiltonian are rarely
available. If we compute the orbital frequency spectrum of an orbit in
an arbitrary coordinate system e.g. in Cartesian or cylindrical
coordinates the resulting orbital frequencies $\Omega_x,\Omega_y,
\Omega_z$ are in no way ``fundamental'' to the nature of the
potential. However as shown by \citep{laskar_90} any canonically
conjugate pair of variables can be combined to obtain a frequency
spectrum, when there are no global actions. These frequencies are
still referred to as ``fundamental frequencies'' since for any regular
orbit all the components of the frequency spectrum are linear integer
combinations of the 3 fundamental e.g. ($\Omega_x,\Omega_y, \Omega_z$)
or ($\Omega_R,\Omega_\phi, \Omega_z$), depending on the coordinate
system selected.

\citet{laskar_90, laskar_93} developed a very accurate numerical
technique ``Numerical Analysis of Fundamental Frequencies'' (NAFF) to
recover frequencies in completely general potentials.  We use an
implementation of this algorithm due to \citet{valluri_merritt_98}
which was adapted for application to orbits in $N$-body potentials by
V10.

The NAFF algorithm recovers orbital frequencies from 3 complex time
series consisting of pairs of phase space variables. For triaxial
potentials we use a Cartesian coordinate system centered on the center
of the galaxy and oriented such that $x,y,z$ correspond to the major
(long), intermediate and minor (short) axes of the potential,
respectively. In the Cartesian coordinates the Fourier analysis is
performed on 3 time series of the form $f_\alpha(t)= \alpha(t)+i
v_\alpha(t)$ (where $\alpha = x, y, z$).

For potentials which are axisymmetric or nearly so, most of the orbits
are tubes which circulate about the symmetry axis. In such potentials
it is preferable to work in cylindrical polar coordinates. We
transform from the planar coordinates $x, y, v_x, v_y$ to plane polar
coordinates $R, v_R, \phi, \Theta$, where $R=\sqrt{x^2+y^2}$,
azimuthal angle $\phi = \arctan({x/y})$, $v_R = (x v_x+y v_y)/R$ and
$\Theta=x v_y - y v_x$. $R, v_R$ are the canonically conjugate radial
coordinate and momentum and hence can be used to define a complex time
series $f_R(t) = R(t)+iv_R(t)$. However, since $\phi$ and $\Theta$ are
the angular coordinate and momentum (and not linear coordinate and
momentum) this pair cannot be used to construct the complex time
series used by the frequency analysis method.  Following
\citet{pap_las96} we use Poincar\'e's symplectic polar variables
$\sqrt{2\Theta}\cos\phi$ and $\sqrt{2\Theta}\sin\phi$, to define the
function $f_\phi = \sqrt{2\Theta} (\cos\phi+i\sin\phi)$. For motion
perpendicular to the equatorial plane we use the complex series
$f_z(t) = z(t)+iv_z(t)$.

V10 demonstrate that fundamental frequencies of orbits in a
self-consistent DF can be used to construct a ``frequency map'' 
which gives a picture of the phase space structure based on its
orbital content.  A frequency map of phase space is obtained by
plotting the {\it ratios} of fundamental frequencies (e.g. in
Cartesian coordinates: $\Omega_x/\Omega_z$ vs. $\Omega_y/\Omega_z$)
for a very large number of orbits.

As we noted above the coordinate system selected to integrate the
orbits and compute the frequencies is determined more by convenience
and convention than any fundamental property of the orbital
frequencies obtained. Nonetheless we will see that for triaxial
systems, the choice of a Cartesian coordinate system aligned such that
the global principal axes of the model coincide with the coordinate
system, result in different orbit families appearing in distinct
groups or lines on a frequency map. The use of an appropriate
coordinate system also allows one to identify truly resonant orbits
families.

Resonant orbits are regular orbits that have fewer than 3 linearly
independent fundamental frequencies which are related via integer
linear equation such as: $l\Omega_1+m\Omega_2+n\Omega_3 = 0,$ where
($l,m,n$) are small integers.  A frequency map can be used to easily
identify the most important resonant orbit families, since such orbits
populate straight lines on such a map. The strength (or importance) of
a resonance can be assessed from the number of orbits associated with
a particular resonance.

\citet{merritt_valluri_99} showed that perfectly resonant orbits in
3-dimensional potentials have two non-zero fundamental frequencies and
occupy thin two-dimensional surfaces (generally multiply connected),
in configuration (physical) space. They are surrounded by a resonance
region consisting of orbits which share the oscillation frequencies of
the perfectly thin resonant parent, but have a third non-zero
frequency, which is small but increases as the orbit deviates from its
resonant parent. Most of these slightly non-resonant orbits also
appear along the resonance lines in the frequency map. Unstable
resonances appear as blank lines or blank spaces on the frequency map.

\begin{figure*}
\centering
\includegraphics[trim=0.pt 0.pt 0.pt 0pt,width=0.252\textwidth, angle=-90]{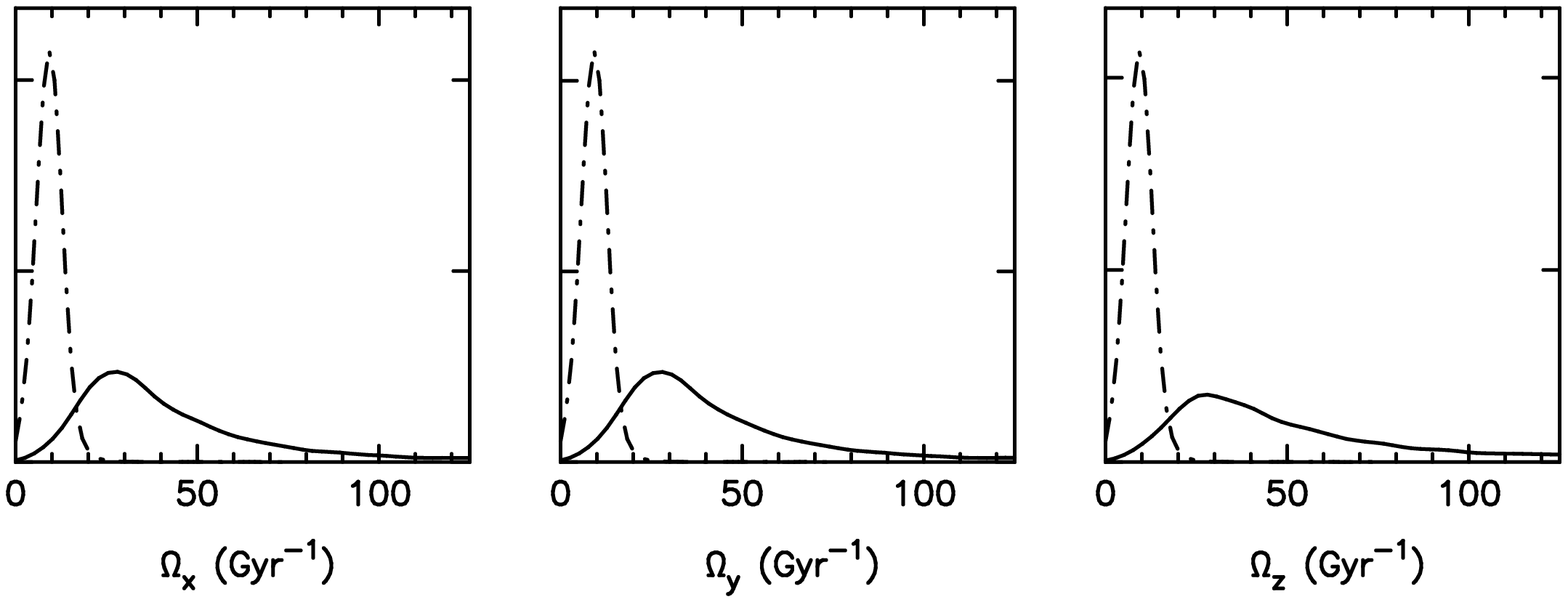}
\includegraphics[trim=0.pt 0.pt 0.pt 0pt,width=0.25\textwidth, angle=-90]{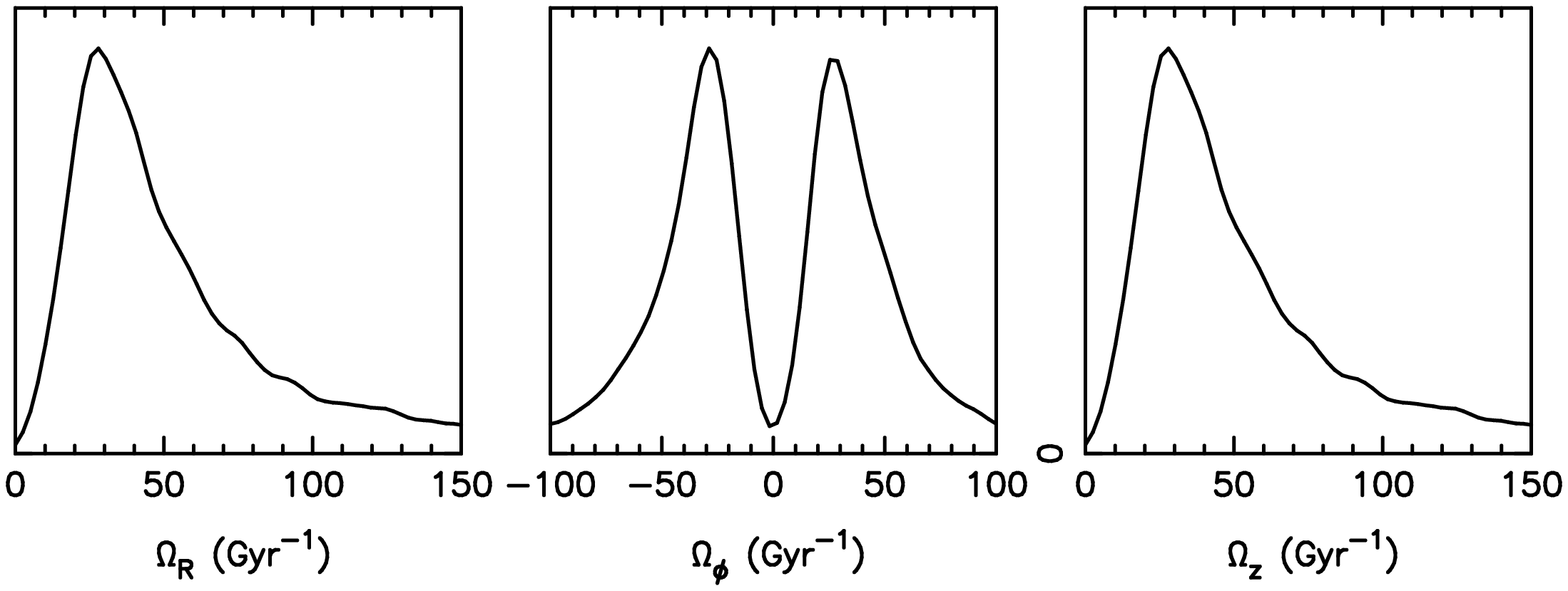}
\caption{Histograms of orbital frequencies of  $10^4$ halo particles with $r_g< 200$~kpc.  The vertical axes have arbitrary scales. Top row (L-R): histograms of $\Omega_x, \Omega_y, \Omega_z$ in a
spherical isotropic NFW halo (dot-dash curves), and after a thin collisionless stellar
disk was grown adiabatically (solid curves); Bottom row (L-R): $\Omega_R$, $\Omega_\phi$, and $\Omega_z$ for the case when a disk is present. The distribution in $\Omega_\phi$ is bimodal because the original halo had
a spherical isotropic DF.}
\label{fig:SphNFW_freqhist}
\end{figure*}

\citet{laskar_etal_92} also showed that since the frequencies of
regular orbits can be recovered with very high accuracy, chaotic
orbits can be easily identified, since their frequencies do not remain
constant but drift when computed over two adjacent time
intervals. They showed that the rate at which the orbits diffuse in
frequency space is correlated with their degree of stochasticity. V10
showed that this way of measuring stochasticity was particularly
useful in $N$-body potentials (and superior to the better known
``Lyapunov exponent'') since it is able to distinguish between
diffusion due to micro-chaos that arises due to discreteness noise
\citep{kandrup_sideris_03} and genuinely irregular behavior. We refer
the reader to V10 for more details.  For each orbit we divide the
integration time of 50~Gyr into two consecutive segments ($t_1$ and
$t_2$) and use NAFF to compute the fundamental frequencies
$\Omega_\alpha(t_1), \Omega_\alpha(t_2)$.  The ``diffusion'' rate for
each frequency component is then computed as:
\begin{eqnarray}
\log(\Delta f_\alpha) = \log{\vert{\frac{\Omega_\alpha(t_1)-\Omega_\alpha(t_2)}{\Omega_\alpha(t_1)}}\vert}.
\label{eq:diff}
\end{eqnarray}
We define the diffusion rate for an orbit, $\log(\Delta f)$ (logarithm
to base 10) to be the value associated with the frequency component
$\Omega_\alpha$ with the highest amplitude $A_\alpha$ measured over
the entire time interval ($t_1+t_2$) \footnote{Note: this definition
of ``diffusion rate'' differs slightly from V10 who used the value
associated with the largest fundamental frequency. The new definition
was found to more accurately identify chaotic orbits and yields a
lower rate of misclassification of regular orbits as chaotic.}. The
larger the value of the diffusion rate $\log(\Delta f)$, the more
chaotic the orbit.  We use the diffusion rate to distinguish between
regular and chaotic orbits, and to distinguish weakly chaotic orbits
from strongly chaotic orbits. It is important to note that for most
systems there is a continuous and nearly Gaussian distribution of
diffusion rates (V10).

Laskar's method recovers orbital frequencies of regular orbits with
very high accuracy in $\sim 20-30$ orbital periods making this method
particularly valuable for studying Galactic halo stars. We integrated
all orbits for 50~Gyr, but only present results for those with orbital
periods shorter than 2.5~Gyr (i.e. those that execute more than 20
orbital periods). All frequencies in this paper are reported
in units of Gyr$^{-1}$.


\section{Results of controlled simulations}
\label{sec:fmaps}

\subsection{Nearly oblate axisymmetric halos}
\label{sec:axisym_maps}
 
\begin{figure*}
\centering
\includegraphics[trim=0.pt 0.pt 0.pt 0pt,width=0.4\textwidth]{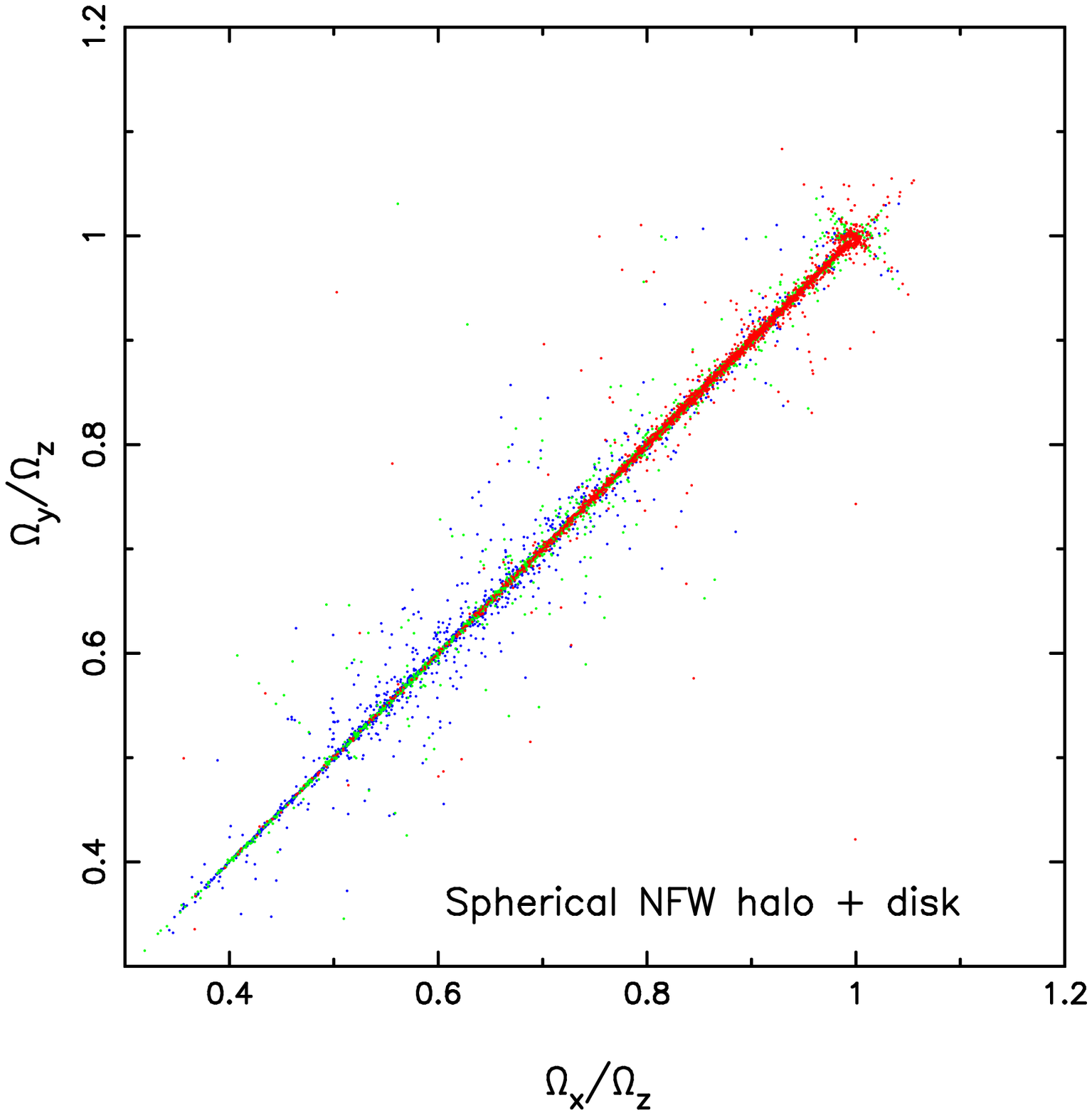}
\includegraphics[trim=0.pt 0.pt 0.pt 0pt,width=0.39\textwidth]{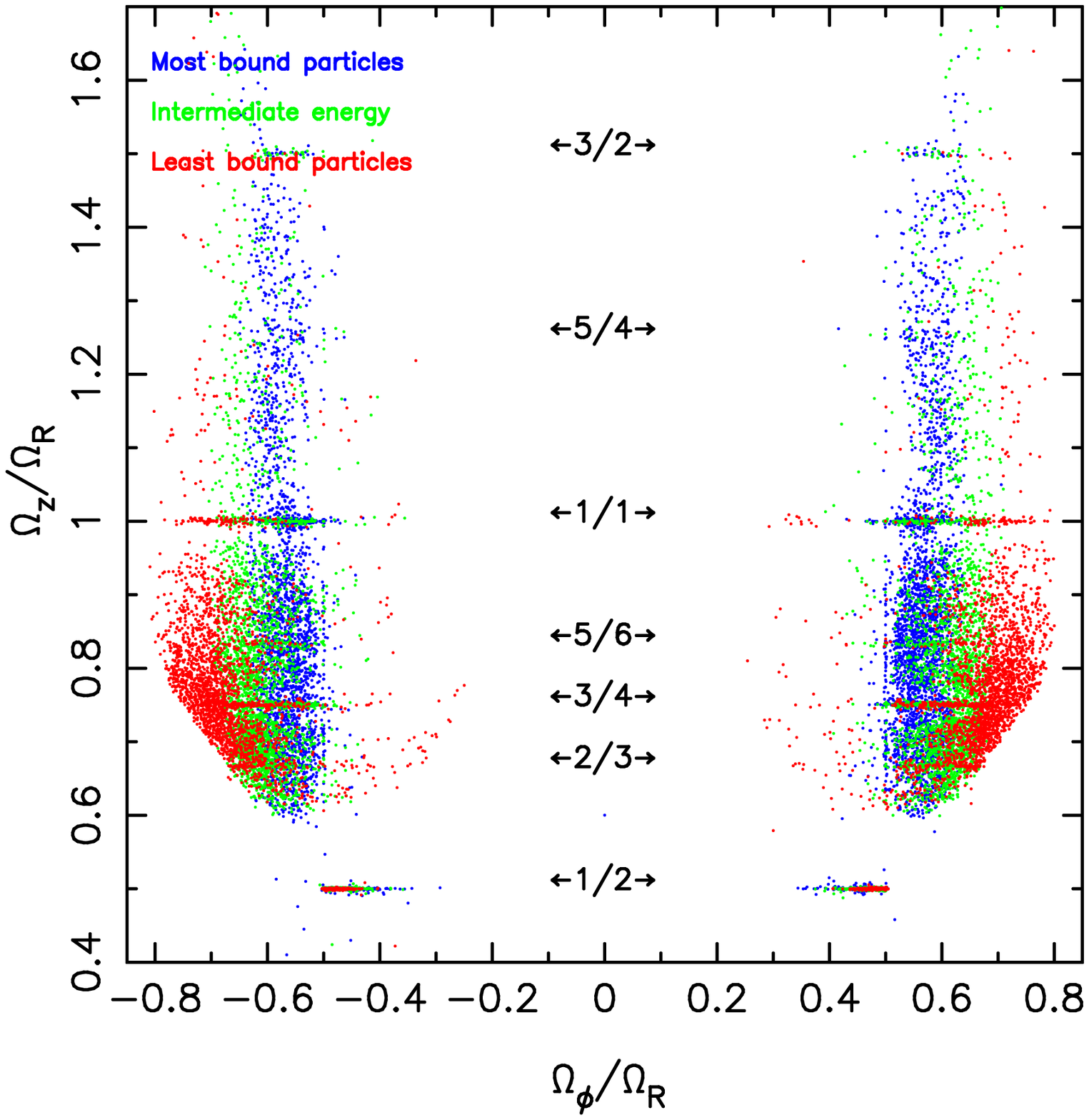}
\caption{Frequency maps of $10^4$ halo orbits in the model with a
stellar disk grown in a spherical NFW halo. Left: in Cartesian
coordinates the map shows that most orbits lie along the diagonal
``resonance line''. This is not a true resonance but represents all
orbits associated with the short-axis tube family.  Right: Frequency
map of the same particles in cylindrical coordinates.  The bisymmetry
about $\Omega_\phi/\Omega_R=0$ results because the halo has no net
rotation. The map shows several resonances which appear as horizontal
lines: e.g. $\Omega_z/\Omega_R=0.5, 0.66, 0.75, 0.83, 1, 1.5$. In both
panels (and hereafter) particles are color coded by binding energy in
three energy bins, each containing 1/3rd of the particles. }
\vspace{4pt}
\label{fig:SphNFW_map}
\end{figure*}

In oblate axisymmetric potentials most orbits are short axis tubes,
hence orbits are best studied in cylindrical coordinates. We begin
with the study of the simplest halo DF: a spherical NFW halo in which
a stellar disk was grown adiabatically (model SNFWD in
Tab.~\ref{tab:simulations}).  Figure~\ref{fig:SphNFW_freqhist} shows
histograms of frequencies for $2\times 10^4$ halo particles with
initial $r_g< 200$~kpc, in Cartesian coordinates $\Omega_x, \Omega_y,
\Omega_z$ (top panels) and in cylindrical coordinates about the
$z$-axis $\Omega_R$, $\Omega_\phi$ $\Omega_z$ (bottom panels). The
initial DF was spherical and isotropic and hence the frequency
distributions in the three Cartesian directions are identical
(dot-dash curves). The growth of an axisymmetric disk (rotating about
the $z$-axis) increases the depth of the potential and makes the halo
flatter in the inner regions. This results in an increase in all the
orbital frequencies, but axisymmetry implies identical distributions
for $\Omega_x$ and $\Omega_y$.  Since the disk potential is
significantly flatter in the $z$ direction, there is a greater
increase in $\Omega_z$ for particles that lie closer to the center of
potential (highest values of $\Omega_z$), accounting for the slight
increase in the weight of the high frequency tail of the $\Omega_z$
distribution.  In cylindrical coordinates the frequency $\Omega_\phi$
(bottom row, middle column) describes the motion in the azimuthal
direction and is either positive or negative depending on whether the
orbit rotates counter-clockwise or clockwise, about the
$z$-axis. Since the halo was set up to be nonrotating, $\Omega_\phi$
values are symmetrically distributed about zero\footnote{Note that the
  frequency $\Omega_R$ is measured relative to the center of the
  cylindrical coordinate system and not relative to a circular orbit
  as in the case of the epicyclic frequency for disk stars.}.

Figure~\ref{fig:SphNFW_map} shows frequency maps for the same
$2\times10^4$ halo particles in Cartesian coordinates (left) and in
cylindrical coordinates (right). Each particle is represented by a
single point whose location is determined by the ratio of the
fundamental frequencies in Cartesian coordinates ($\Omega_x/\Omega_z$
vs. $\Omega_y/\Omega_z$) or cylindrical coordinates
($\Omega_z/\Omega_R$ vs.  $\Omega_\phi/\Omega_R$). In frequency maps
the color of a point represents the binding energy of its
orbit with blue representing the 1/3rd most tightly bound
particles in the map, red representing the 1/3rd least bound particles
and green representing the intermediate energy range.
 
Since the growth of the disk makes the originally spherical
distribution of particles oblate axisymmetric, the DF is entirely
populated by short-axis tube orbits. In a Cartesian frequency map
(Figure~\ref{fig:SphNFW_map} left panel) such orbits primarily lie
along a diagonal line that satisfies the condition $\Omega_x/\Omega_z
\sim \Omega_y/\Omega_z$. Most short axis tubes are not ``resonant''
orbits, but in a Cartesian frequency map they all appear clustered
along a line, because each short-axis tube can be viewed as arising
from a radial perturbation of a parent ``thin shell'' orbit which is a
resonant orbit \citep{dezeeuw_hunter_90}.

Truly resonant short-axis tube orbits are those that appear clustered
along lines in a frequency map in cylindrical coordinates
(Figure~\ref{fig:SphNFW_map} right panel). This map shows bisymmetry
between the right and left halves, reflecting the bimodal distribution
in $\Omega_\phi$ (Fig.~\ref{fig:SphNFW_freqhist}).  The frequency map
in cylindrical coordinates shows a striking number of resonances which
appear (primarily) as horizontal lines delineated by the enhanced
clustering of particles at resonances between the vertical oscillation
frequency $\Omega_z$ and radial oscillation frequency
$\Omega_R$. Resonances are seen at $\Omega_z/\Omega_R = 0.5, 0.66,
0.75, 0.83, 1, 1.5$ (i.e. $\Omega_z/\Omega_R= 1/2; 2/3, 3/4, 5/6,
1/1,3/2$), as indicated by labels.  Thus the growth of a disk in a
spherical potential results in halo orbits becoming resonantly trapped
at numerous resonances, primarily between the radial and vertical
oscillation frequencies.

\subsection{Triaxial halos with rigid disks}
\label{sec:triax_maps_rigid}

\begin{figure}
\vspace{0.4cm}
\centering
\includegraphics[trim=0.pt 0.pt 0.pt 0pt,angle=-90.,width=0.4\textwidth]{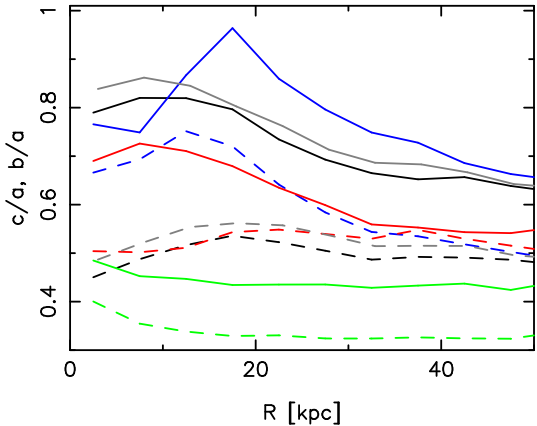}
\caption{Halo axis ratios $b/a$ (solid curves) and $c/a$ (dashed
  curves) as a function of radius. The green curves show the shape of
  the initial triaxial halo. The black curves are for the triaxial
  halo + short axis disk (model SA1), the blue curves are for the
  triaxial halo + long axis disk (model LA1), the red curves are for
  triaxial halo + intermediate axis disk (model IA1) and the grey
  curves are for the triaxial halo + tilted disk (model TA1).}
\label{fig:shape_param}
\end{figure}

We now consider controlled simulations of galactic potentials with
triaxial halos. The simulations are taken from D08.  We consider four
different orientations of the disk relative to the same triaxial halo
(Halo A from D08).  In all four cases the disk potential consisted of
particles, but the particles were held rigid while the halo was
allowed to relax. In all four cases, the inner region of the halo
became less triaxial (more oblate) with short axis aligned with the
symmetry axis of the disk. We define $a,b,c$ to be the density
semi-axes of the major, intermediate and short axis respectively of
the global halo. 
 
The shapes of the triaxial halos were measured as described  in D08. Briefly, 
we measure the eigenvalues of the unweighted moment of inertia tensor $I$ obtained  in bins of  $N$ particles \citep[][see also eq.  2 from D08]{katz_91}. The ratios of the eigenvalues of the diagonalized moment of inertia tensor   ($\mathcal I_{11}> \mathcal I_{22}> \mathcal I_{33}$) are used to calculate the axis ratios $b/a$ and $c/a$. The shapes were measured using an iterative procedure in annular shells of fixed semi-major axis width and are differential rather than integrated over all particles within a given ellipsoidal radius.

\begin{figure*}
\vspace{0.4cm}
\centering
\includegraphics[trim=0.pt 0.pt 0.pt 0pt,angle=0.,width=0.22\textwidth]{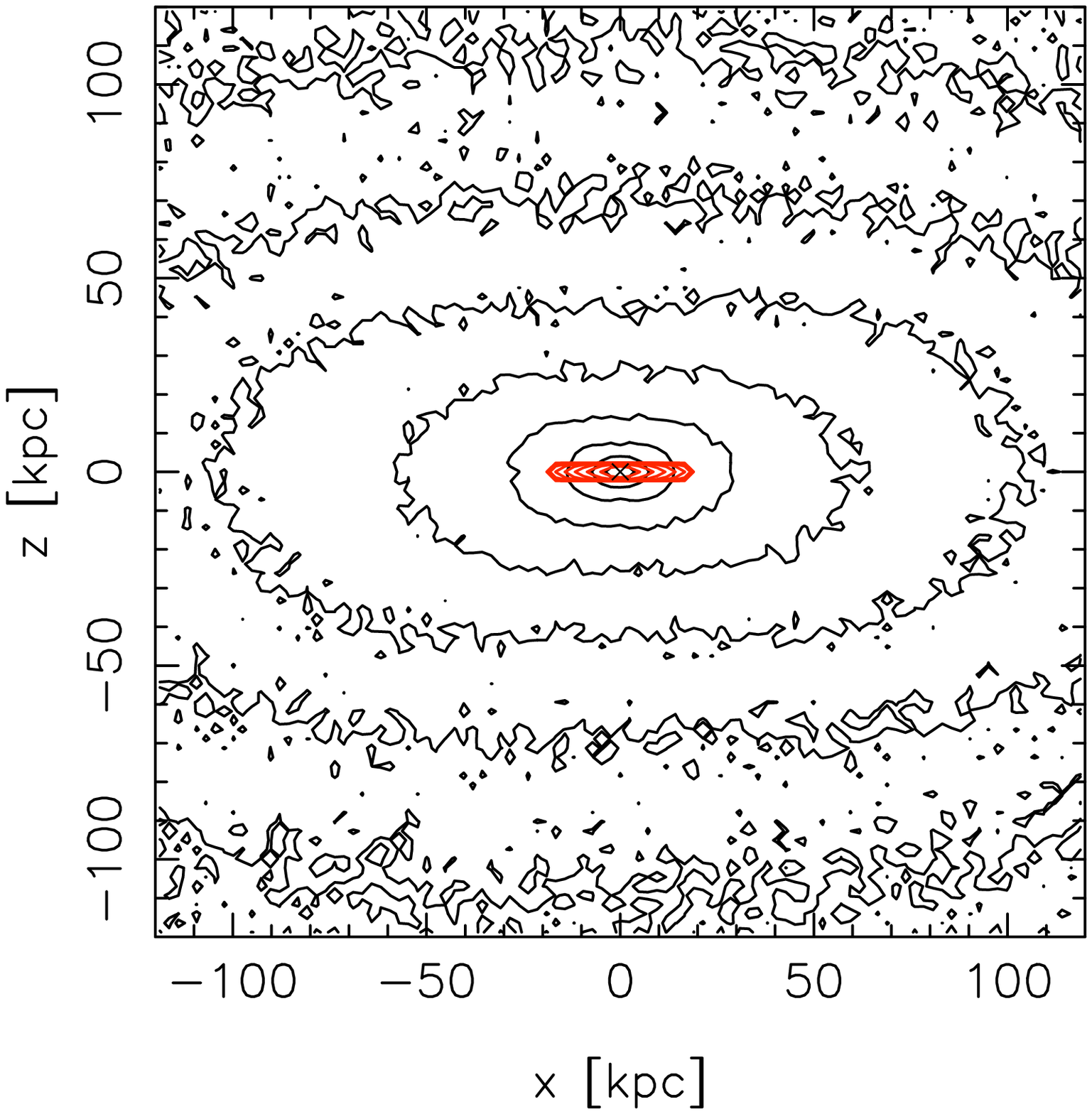}
\includegraphics[trim=0.pt 0.pt 0.pt 0pt,angle=0.,width=0.22\textwidth]{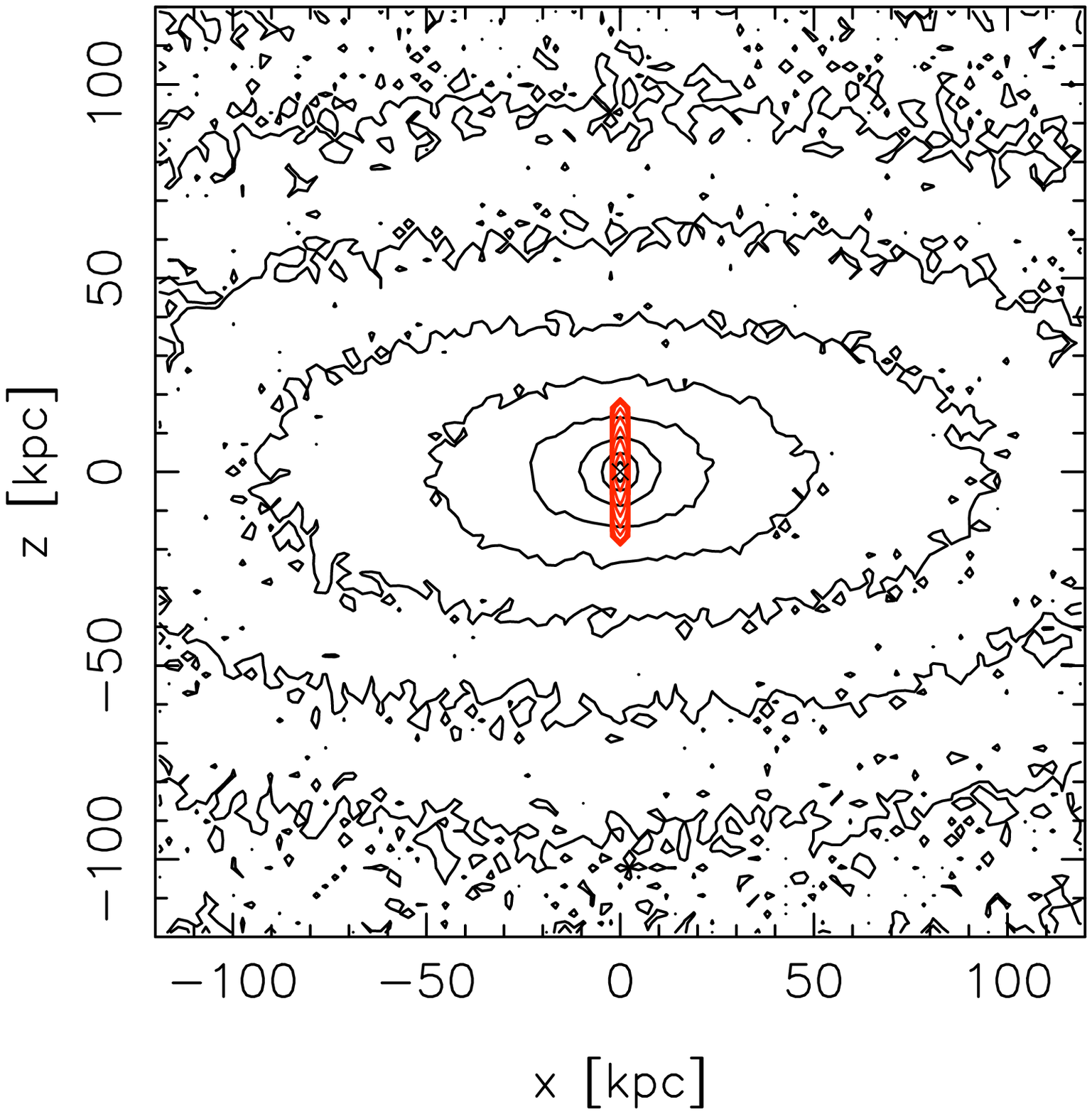}
\includegraphics[trim=0.pt 0.pt 0.pt 0pt,angle=0.,width=0.22\textwidth]{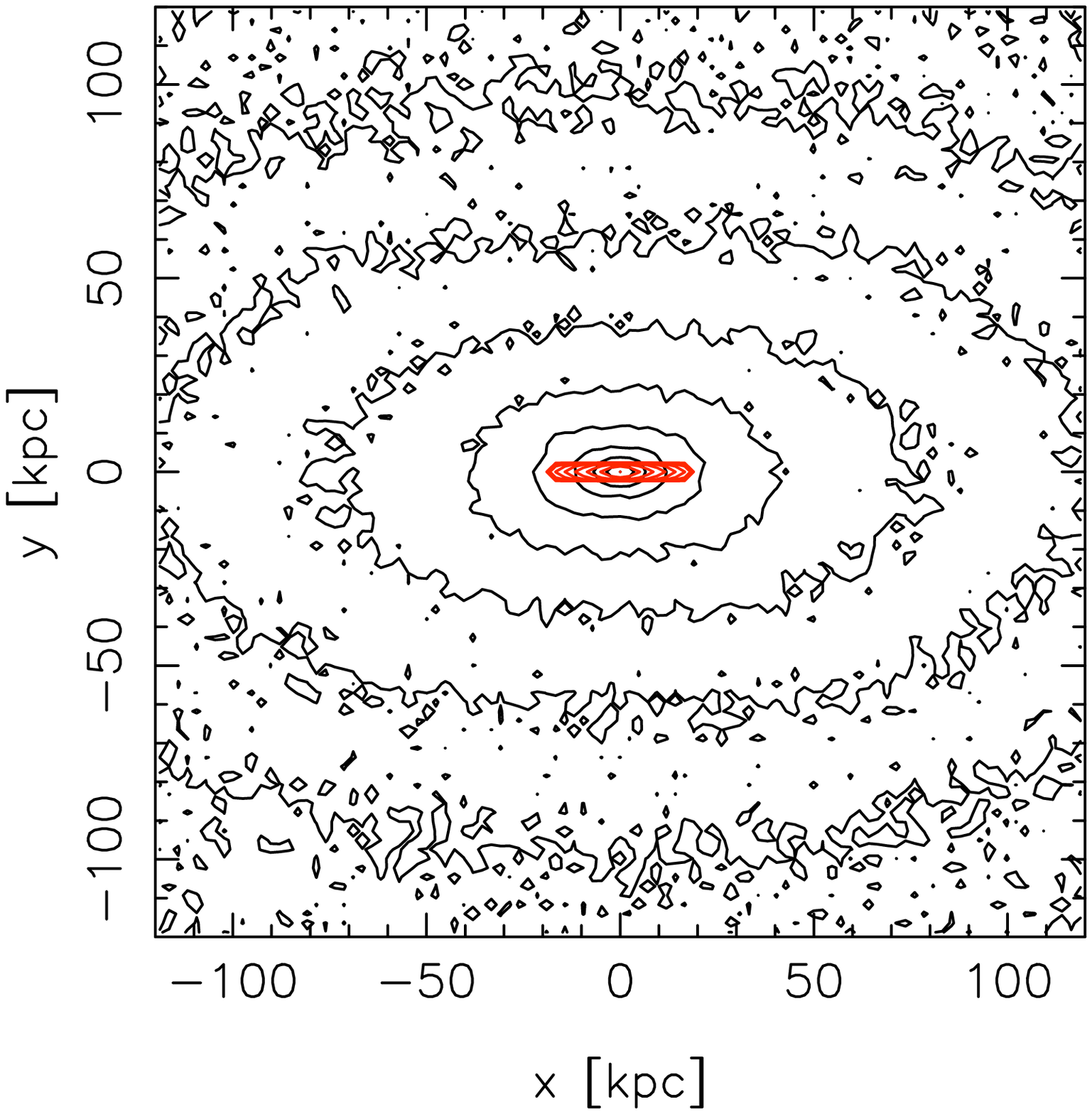}
\includegraphics[trim=0.pt 0.pt 0.pt 0pt,angle=0.,width=0.22\textwidth]{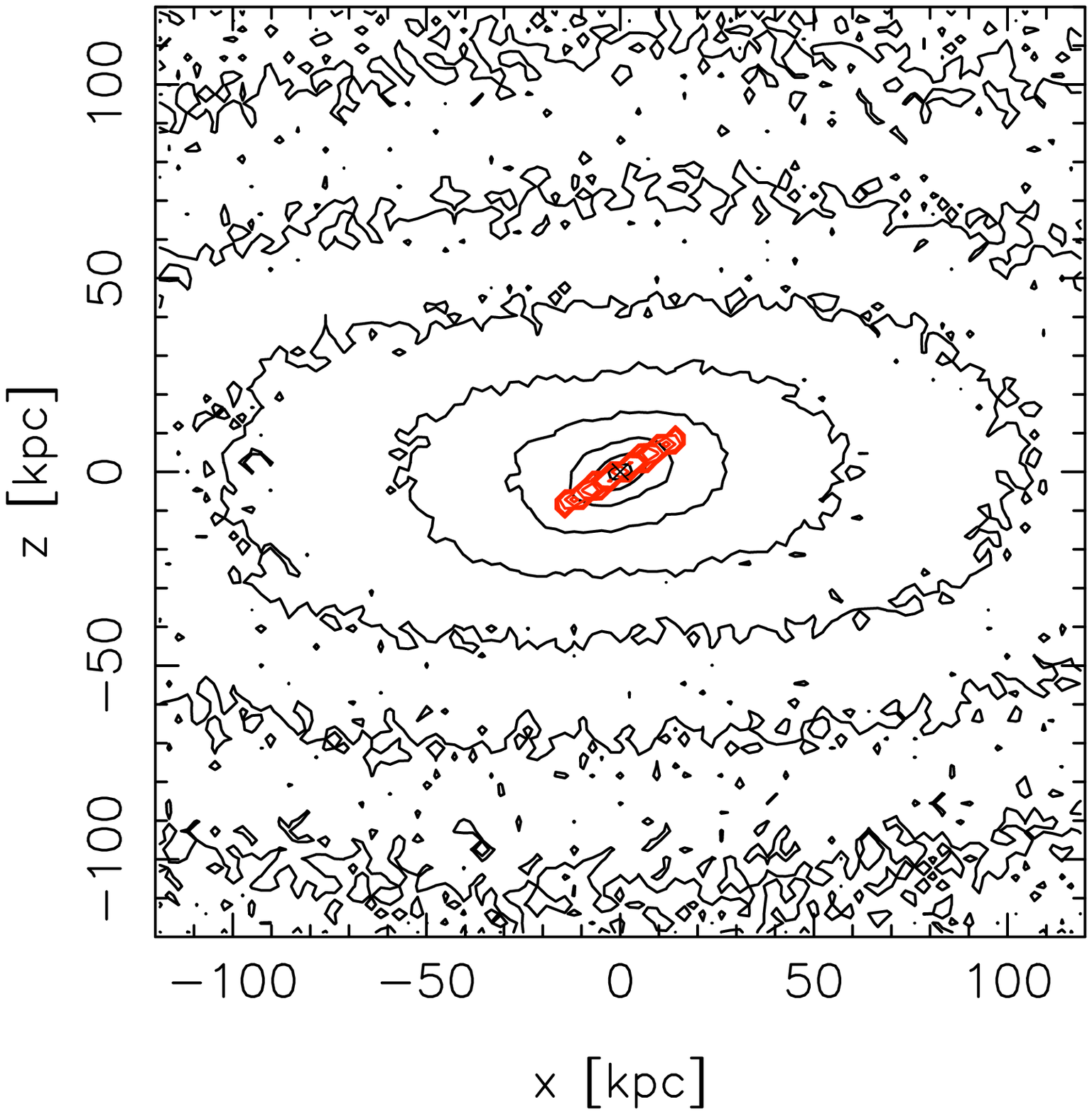}
\includegraphics[trim=0.pt 0.pt 0.pt 0pt,angle=0.,width=0.22\textwidth]{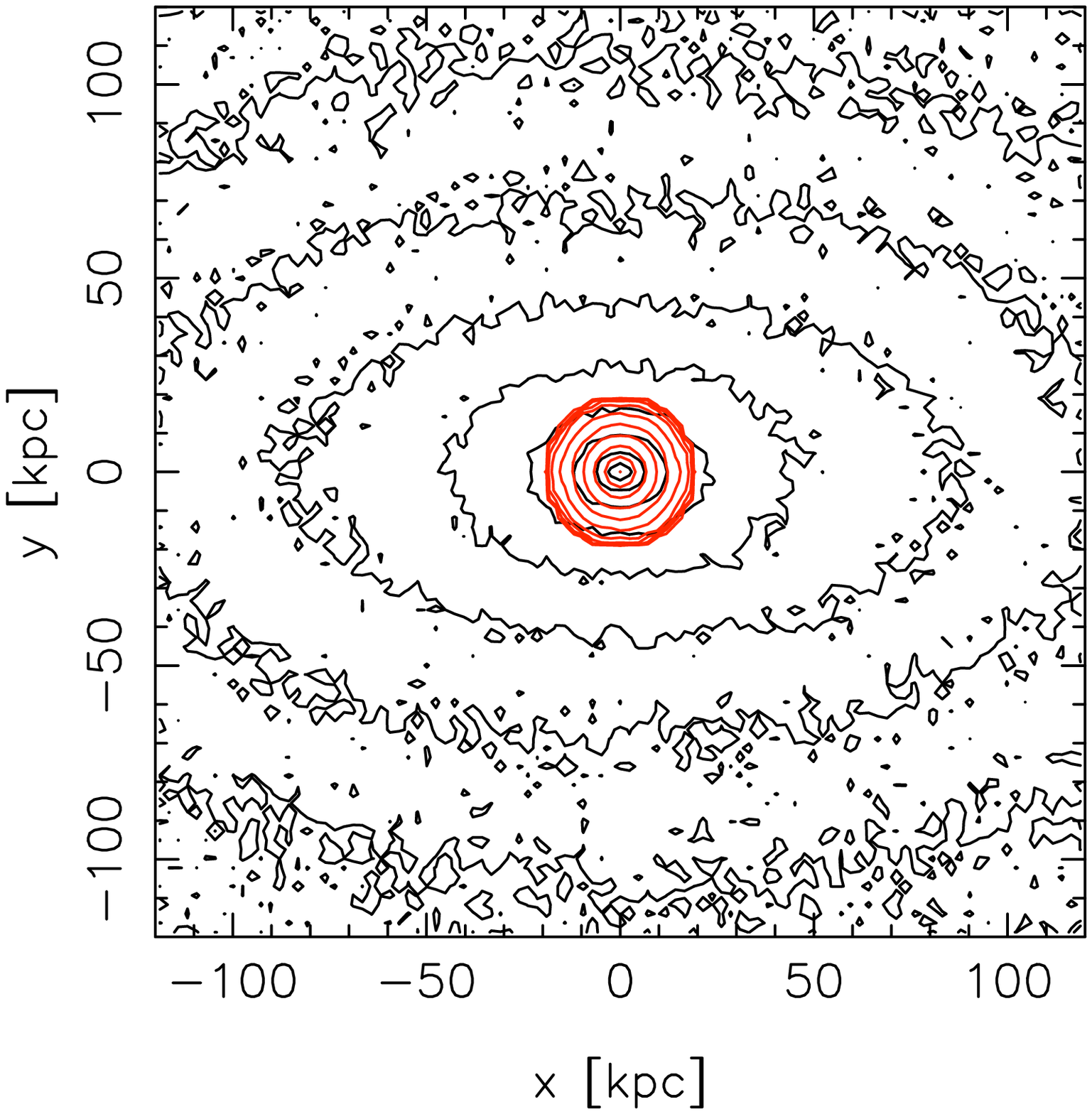}
\includegraphics[trim=0.pt 0.pt 0.pt 0pt,angle=0.,width=0.22\textwidth]{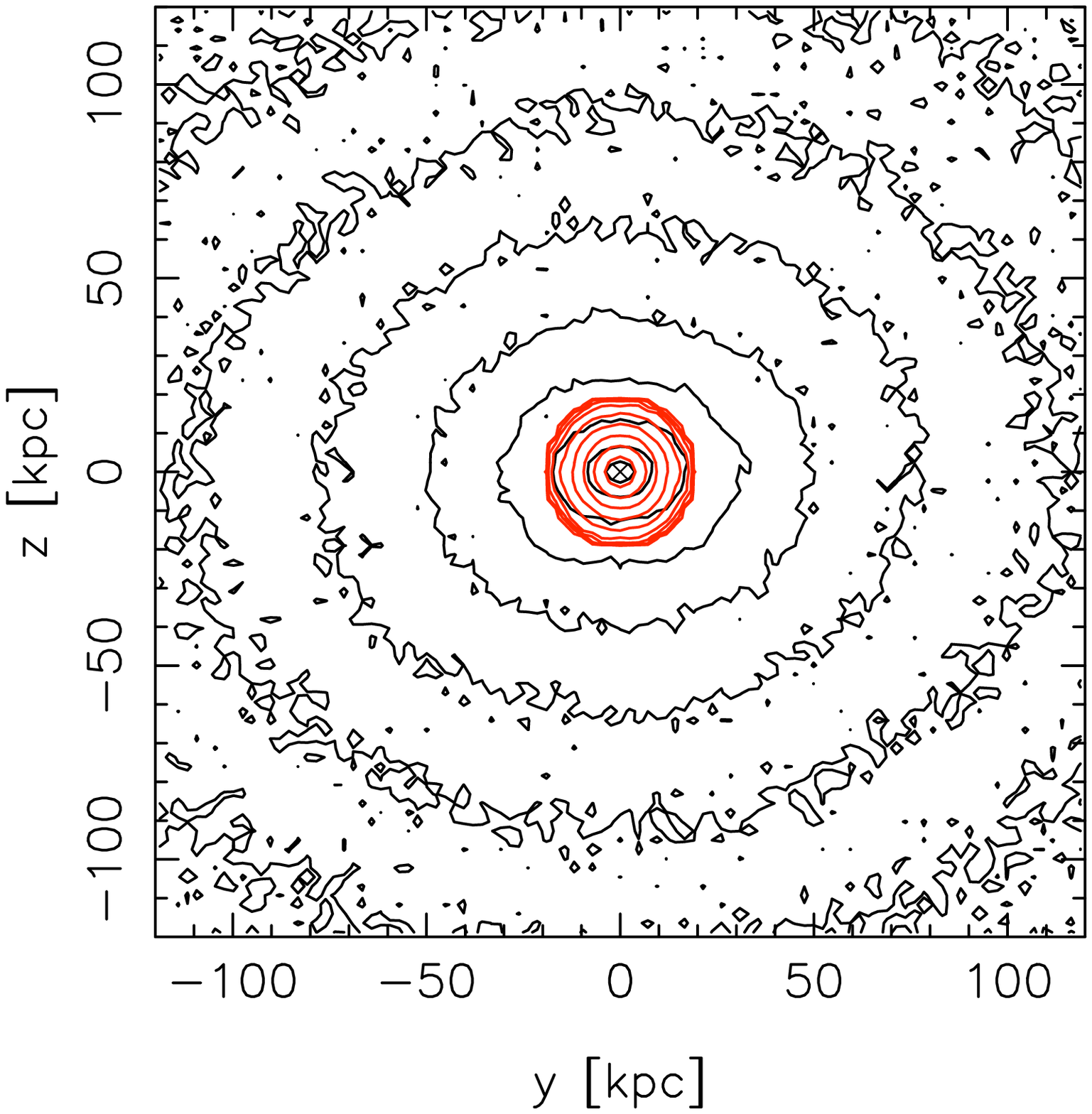}
\includegraphics[trim=0.pt 0.pt 0.pt 0pt,angle=0.,width=0.22\textwidth]{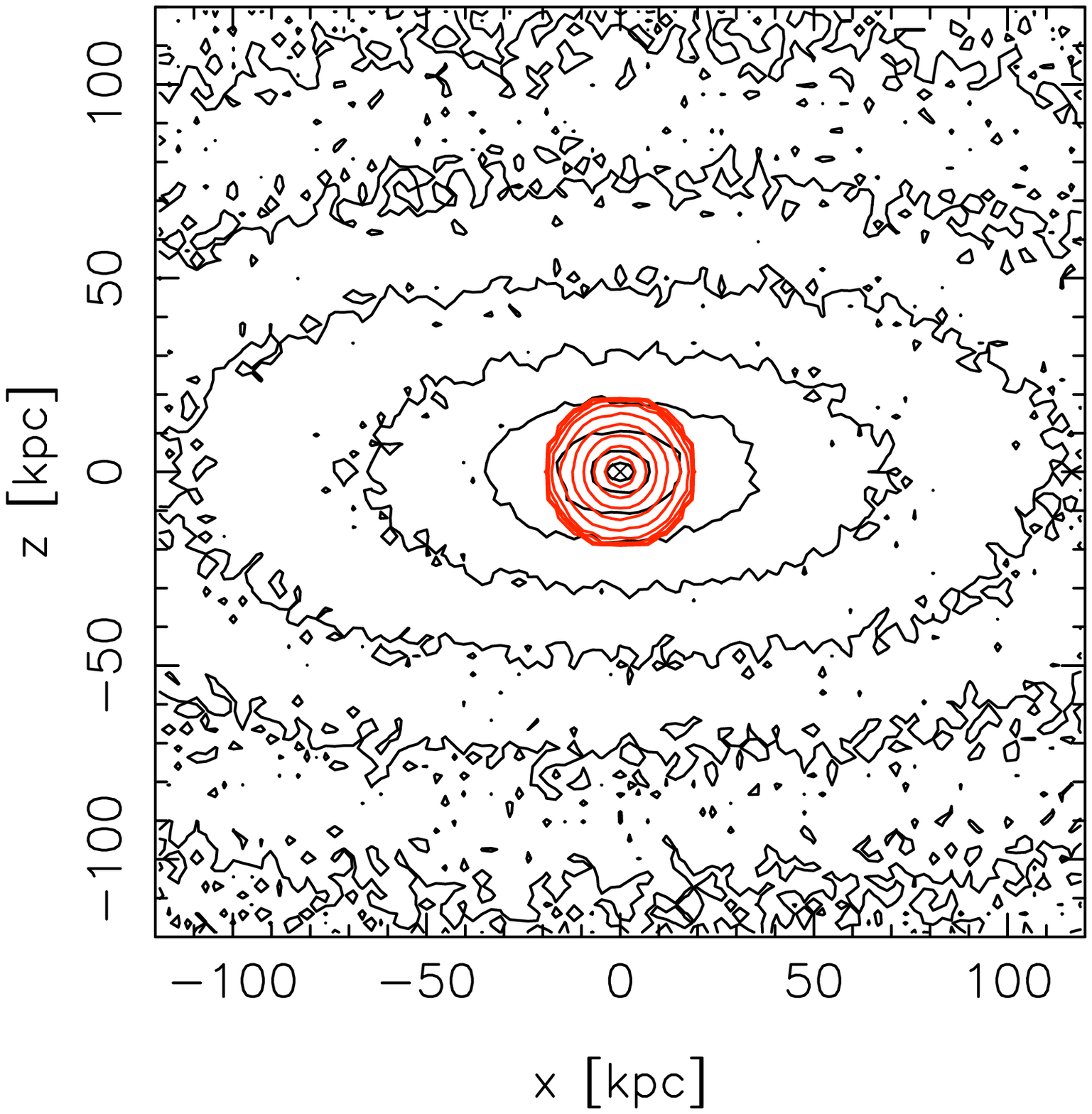}
\includegraphics[trim=0.pt 0.pt 0.pt 0pt,angle=0.,width=0.22\textwidth]{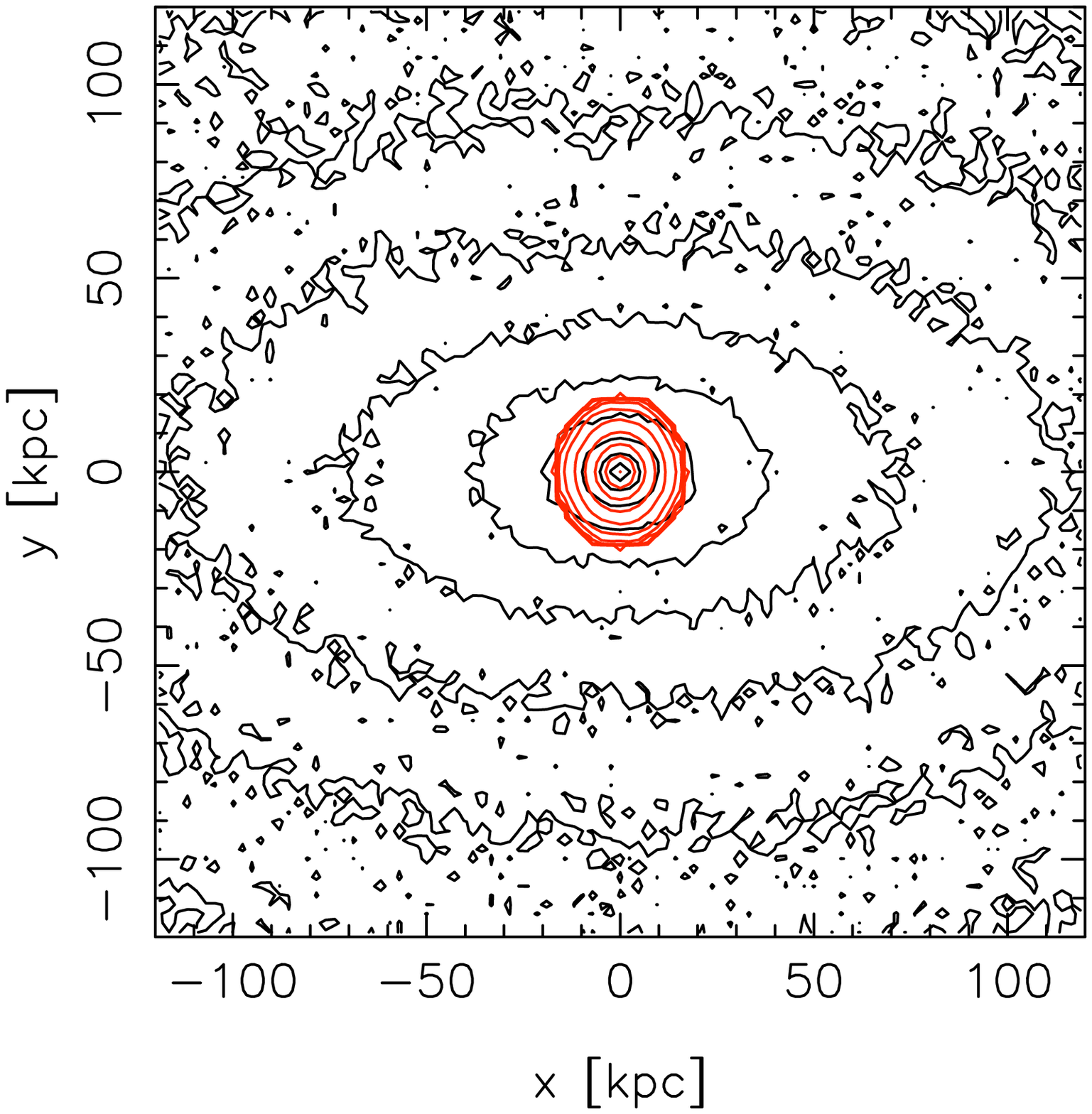}
\caption{Density contours of halo mass distribution in triaxial halos
  after disk (shown by red contours) is grown with spin axis aligned
  in different ways. From Left to right: models SA1, LA1, IA1 and
  TA1. Top row: density contours of halos with disk seen
  edge-on. Bottom row: density contours of halos in the plane of the
  disk. Halo triaxiality varies with distance from the center of the
  potential.}
\label{fig:shape_contours}
\end{figure*}

 Figure~\ref{fig:shape_param} shows the change in the
axes ratios $b/a$ and $c/a$ as a function of radius. The green curves
show that the initial triaxial halo A was very strongly
prolate-triaxial (a result of low angular momentum mergers). The black
curves are for model SA1 (triaxial halo + short axis disk), the blue
curves are for model LA1 (triaxial halo + long axis disk), the red
curves are for model IA1 (triaxial halo + intermediate axis disk), and
the grey curves are for model TA1 (triaxial halo + tilted  disk).
Regardless of the orientation of the disk, we see that its growth
results in a very significant increase in axis ratios $b/a$ (solid
curves) and $c/a$ (dashed
curves) within the inner 30~kpc, for models SA1, LA1, TA1,
and a moderate increase in oblateness within 50~kpc for model
IA1\footnote{Recall that as $b/a \rightarrow 1$, a model becomes more
oblate.}. (D08 shows the change in halo shape out to 200~kpc.) Model
LA1 (blue curves) shows the most significant change in shape over the
radial range $\sim 20-50$~kpc.  In the inner 15~kpc, the model is more
triaxial than it is at larger radii, due to the formation of an inner
elongation along the $y$-axis (see Fig.~\ref{fig:shape_contours}).

Although all the models become more oblate at the center, they still
remain triaxial both at small and large radii. This is seen in
Figure~\ref{fig:shape_contours} which shows contour plots of the dark
matter halo projected density (black curves) in two projections for each of the
models studied. The top row shows density contours of the dark matter
particles when viewed with disk edge-on as represented by the red
contours.  The bottom row shows the density contours of the halo in
the plane of the disk, except in model TA1 (where the disk is inclined to the principle planes). The top panels show that
in all cases the contours become flattened with the symmetry axis
co-aligned with the disk's symmetry axis at radii $<15$~kpc, but
retain their original elongation along the $x$-axis at larger
radii. The lower panels show that although the halo becomes flattened
in the inner region, it is not axisymmetric, even in the plane of the
disk, showing that triaxiality varies with radius.

In the rest of this section we will show frequency maps of orbits in
each of these models. In all cases we show frequency maps in Cartesian
coordinates for $\sim 10^4$ halo orbits selected in two ways. First,
orbits were randomly selected to have an initial distance from the
galactic center $r_g < 200$~kpc. Since these frequency maps represent
a randomly selected subsample of orbits drawn from the entire dark
matter halo we will refer to these maps as showing the ``full halo
DF''.

Ideally we would like to be able to distinguish between different halo
orientations relative to the disk from the DF of halo {\it stars}
since the orbits of dark matter particles are not directly
observable. However, halo stars are known to be significantly more
centrally concentrated than the dark matter
\citep{battaglia_etal_05}. Therefore we also present frequency maps of
orbits selected in a second way, where the orbits satisfy two
conditions: (i) their initial distance from the ``sun'' $R_s <10$~kpc,
(ii) each orbit has an apocenter radius $r_{\rm apo} <50$~kpc from the
galactic center. Restriction (i) is motivated by the expectation that
{\it Gaia} \citep{perryman_etal_01} will obtain 6 phase space
coordinates (and in particular the most accurate distances and proper
motions) for stars within 10~kpc of the sun. Restriction (ii) is
imposed because we do not anticipate that any method of halo shape
determination will provide an accurate measurement of the shape of the
halo at distances greater than 50~kpc from the Galactic center. By
restricting ourselves to only those stars with $r_{\rm apo} <50$~kpc,
we are ensuring that the maps only contain orbits which explore the
region of the halo where the shape is well determined. We will refer
to this second set of maps as showing the ``inner halo DF''.
Figures~\ref{fig:shape_param} \& ~\ref{fig:shape_contours} show that
the innermost region of the halo becomes significantly oblate due to
the presence of the disk, regardless of orientation of the large scale
halo.

In the sections that follow we will compare the DFs of the four model
by showing that the frequency map representations of their DFs differ
significantly from each other regardless of whether the full halo or
inner halo is represented. This is remarkable since the initial DFs of
each halo (prior to the growth of the disk) were identical and
Figure~\ref{fig:shape_contours} shows that while the disk alters the
inner regions of the halo it remains triaxial and elongated along its
original long-axis at large radii.

It is important to note that in all cases we begin with exactly the
same set of $10^4$ orbits from the original triaxial halo (whose
frequency map is plotted in Fig.~\ref{fig:SA1_map} left), and follow
that same set of orbits in the different models in which disks were
grown adiabatically.

\subsubsection{SA1: Triaxial halo with short axis disk}
\label{sec:SA1}

\begin{figure*}
\centering
\includegraphics[trim=0.pt 0.pt 0.pt 0pt,width=0.32\textwidth]{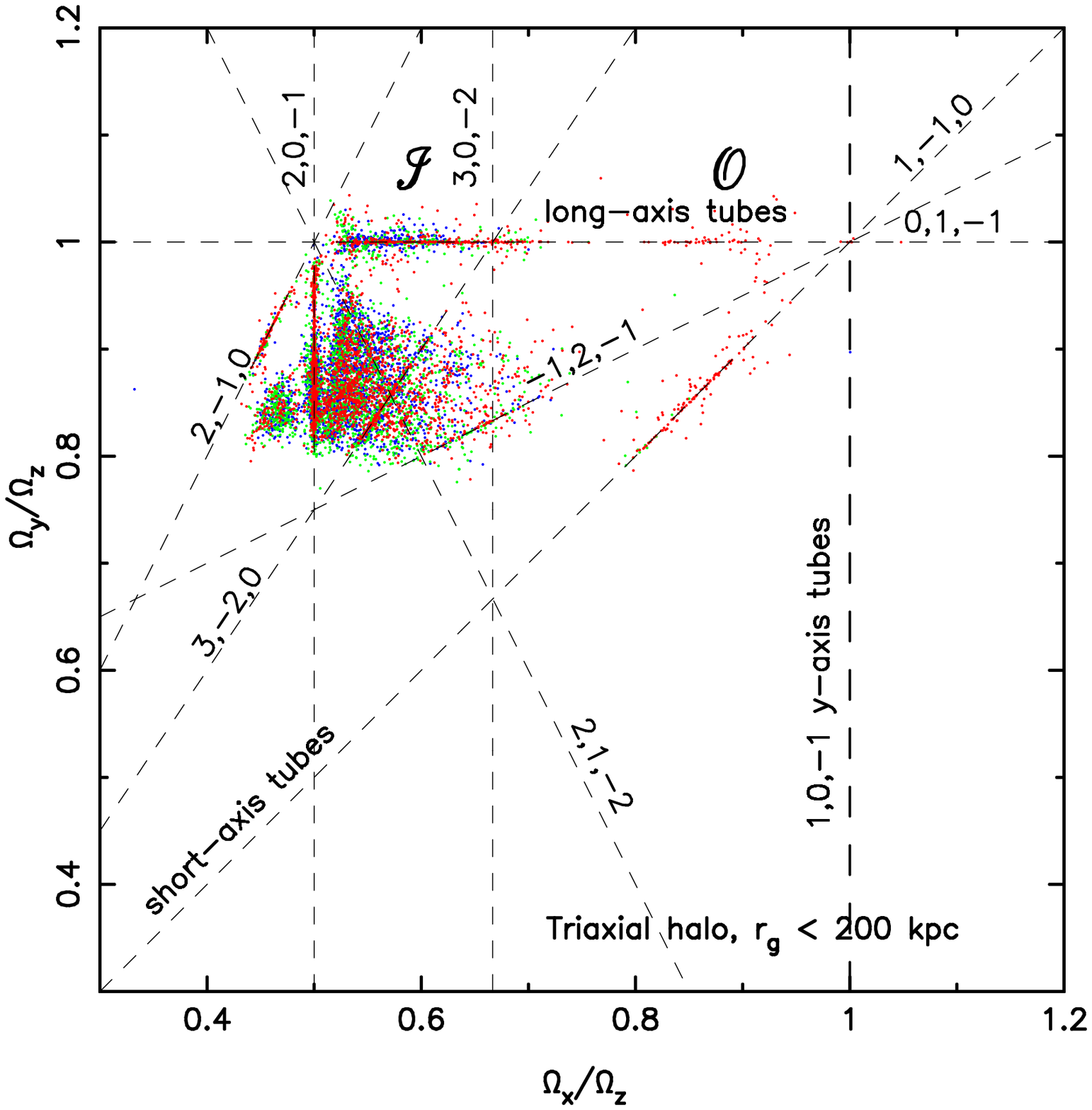}
\includegraphics[trim=0.pt 0.pt 0.pt 0pt,width=0.32\textwidth]{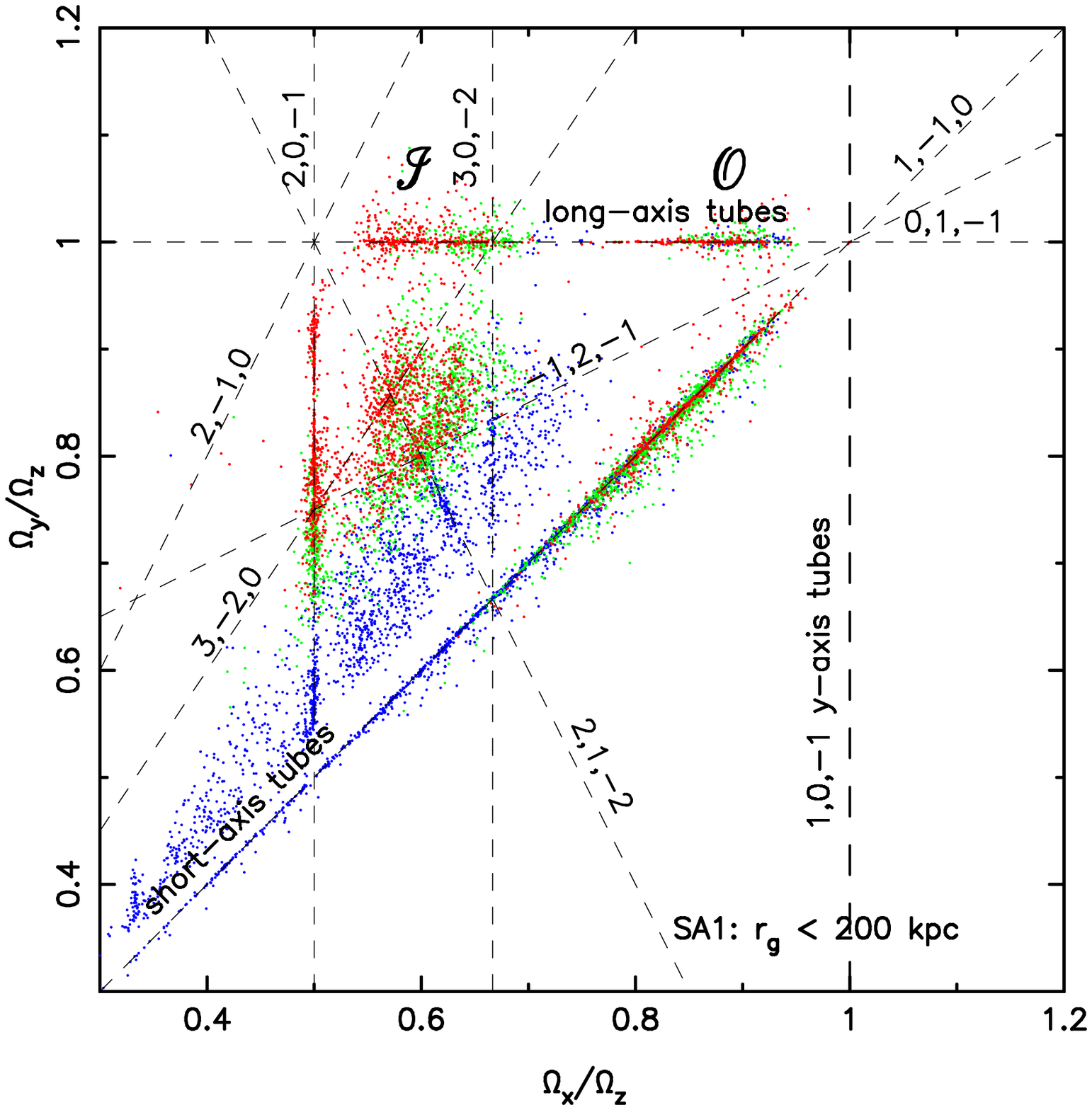} 
\includegraphics[trim=0.pt 0.pt 0.pt 0pt,width=0.32\textwidth]{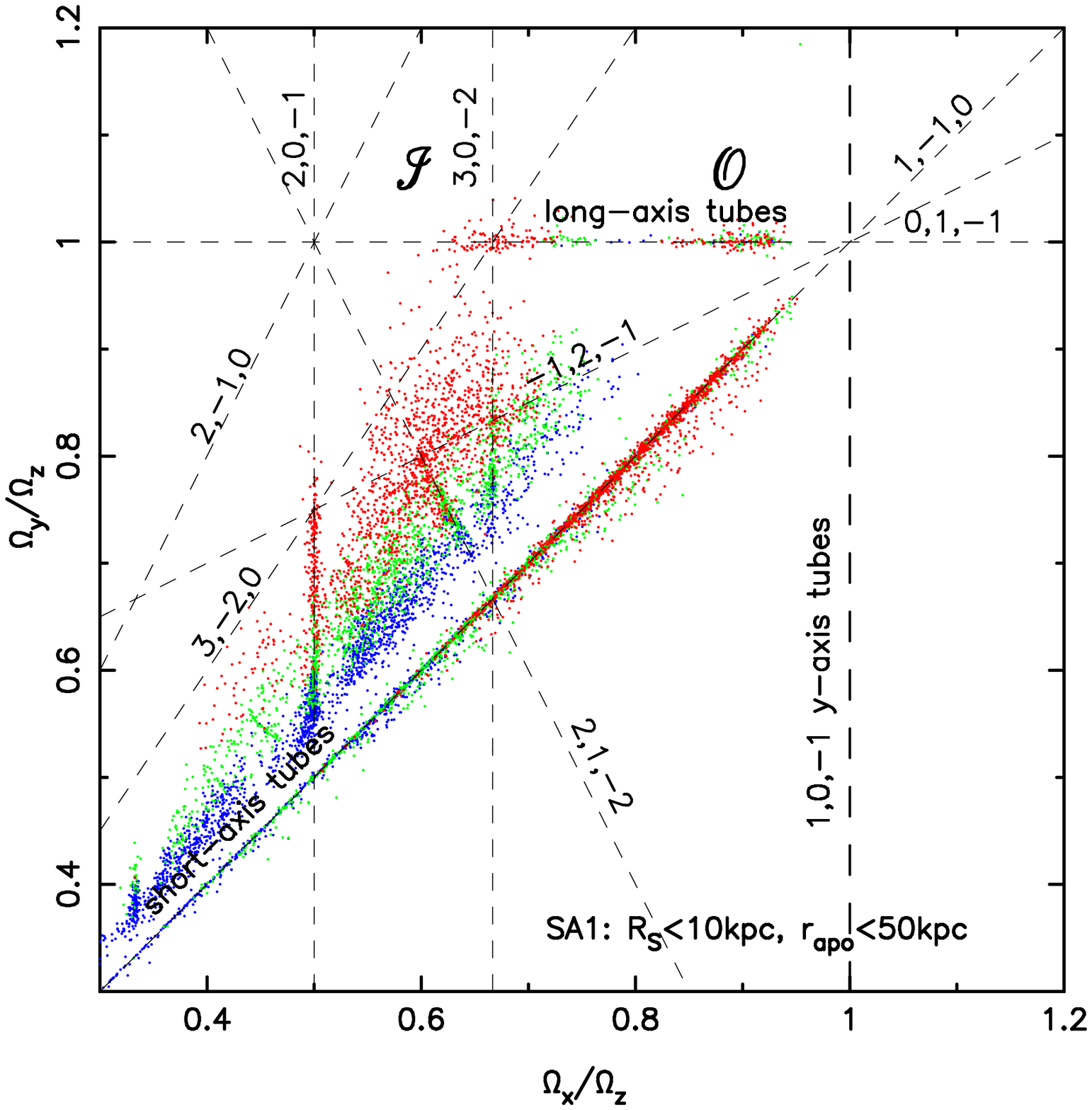}\vspace {0.1in}
\caption{Cartesian frequency maps of $\sim 10^4$ halo particles with
  $r_g < 200$~kpc in the original triaxial halo (left); frequency map
  of the same particles after the growth of a disk perpendicular to
  the short-axis (model SA1) (middle). Right: Cartesian frequency map
  for particles with $R_s<10$~kpc from the ``sun'' and $r_{\rm
    apo}<50$~kpc. In all panels, dashed lines mark important
  resonances, also labeled by resonance integers $(l,m,n)$;
  approximate locations of inner ($\mathscr I$) and outer ($\mathscr
  O$) long-axis tubes along a horizontal line $\Omega_y/\Omega_z \sim
  1$ are marked.  Short-axis tubes are along the diagonal line with
  $\Omega_x/\Omega_y \sim 1$. Particles are color coded by energy (see
  text). }
\label{fig:SA1_map}
\end{figure*}

Figure ~\ref{fig:SA1_map} (left) shows the frequency map of a triaxial
halo that was formed from multiple mergers of spherical NFW halos
(This model is the original triaxial halo in which all the disks are
grown and is referred to as ``Halo A'' in D08 and V10.) The particles
are color coded by energy as described above.

V10 used the relationships between the fundamental frequencies of
orbits in this triaxial potential to characterize them by orbital
type. They showed that 86\% of the particles were on box orbits, 11\%
were on long-axis tube orbits, 2\% were short-axis tubes and 1\% were
chaotic. The fact that box orbits dominate over long-axis tubes and
short axis tubes in this model strongly reflects its formation history
- from a two-stage binary merger, with little angular momentum. Box
orbits do not have any net angular momentum and hence their 3
fundamental frequencies are (in general) uncorrelated with $\Omega_x
\lesssim \Omega_y \lesssim \Omega_z$. This means that box orbits are
generally found scattered in the frequency map at $\Omega_y/\Omega_z <
1$ and to the left of the diagonal with $\Omega_x/\Omega_y <1$. The
long-axis ($x$) tubes have $ \Omega_y \sim \Omega_z$ and hence
primarily cluster along a horizontal line at $
\Omega_y/\Omega_z=1$\footnote{Long axis tubes and short axis tubes
  generally are not considered ``resonant'' orbits, however they may
  be viewed as perturbations of their respective ``thin shell''
  orbits \citep{dezeeuw_hunter_90}.}. The inner long-axis tubes lie at smaller
$\Omega_x/\Omega_z$ values near the label ``{\it$ \mathscr I$}''
(between the two vertical lines corresponding to the resonances
(2,0,-1) and (3,0,-2)) and the outer long-axis tubes lie at larger
$\Omega_y/\Omega_z$ values near the label ``{\it$ \mathscr O$}''. The
fact that this distribution function has only a tiny fraction of short
axis tubes (2\%) as determined by more rigorous orbital classification
in V10, is represented by the sparse distribution of red points along
the diagonal $ \Omega_x/\Omega_y=1$. Thin dashed lines mark several
possible resonance lines, but we see that only the two ``banana
orbit'' families labeled (2, -1, 0) and (2, 0, -1) are prominent. The
qualitative dominance of the box-orbits relative to the long-axis
tubes and short-axis tubes can be visually assessed directly from the
frequency map, without specifically going through the process of orbit
classification as done by V10 and others \citep{carpintero_aguilar_98,
  hoffman_etal_09, hoffman_etal_10}. We will see that this becomes
particularly important for identifying orbits in halos whose shapes
vary with radius.

Figure~\ref{fig:SA1_map} (middle) shows the frequency map of the same
set of halo orbits after the growth of a stellar disk perpendicular to
the short axis of the halo (model SA1).  V10 found that the fraction
of box orbits has now decreased from 86\% in the original triaxial
halo to 48\% after the growth of the disk, but the long-axis tube
fraction remained almost the same (12\%). They found a significant
increase in the fraction of short-axis tubes (2\% to 31\%) and chaotic
orbits (1\% to 9\%). The increased fraction of short-axis tubes
relative to the long-axis tubes and box orbits is obvious in the
frequency map, where the short-axis tubes appear as the enhanced
clustering along the diagonal line.  The map shows that the disk also
causes orbits to become separated into distinct bands in energy
(color), with the most tightly bound (blue) particles migrating to the
bottom left hand corner.

The migration of particles in the frequency map occurs because the
introduction of the disk potential perpendicular to the $z$-axis,
increases the vertical frequency ($\Omega_z$) of orbits more
significantly than either of the other two frequencies. Since the disk
is centrally concentrated, the increase in $\Omega_z$ is particularly
significant for orbits which are deeper in the potential (colored
blue). Therefore both $\Omega_y/\Omega_z$ and $\Omega_x/\Omega_z$
decrease, and the blue points move downward and to the left.  The
growth of a disk also increases the fraction of halo orbits trapped in
resonances, but some resonances are destroyed. For instance the
vertical (2,0,-1) ``banana resonance'' has significantly increased in
prominence (due to trapping of orbits in the $x-z$ plane, but the
(2,-1,0) banana resonance in the triaxial model (which lies in the
$x-y$ plane) is destroyed because this is the disk plane, and the
presence of the disk decreases the degree of triaxiality. New
resonances are also populated e.g. (3,0,-2) the ``fish'' resonance,
and (2,1,-2) resonance\footnote{3-dimensional images of all the major
resonances in triaxial potentials are shown in
\citet{merritt_valluri_99}.}. Since a frequency map represents the
{\it ratios} of the frequencies and not the frequencies themselves, it
is insensitive to the absolute value of the energy of individual
particles and it is therefore possible to identify the {\em global}
orbital families and resonances (i.e. those that are important over a
large range of energies).
 
\begin{figure*}
\centering
\includegraphics[trim=0.pt 0.pt 0.pt 0pt,width=0.35\textwidth]{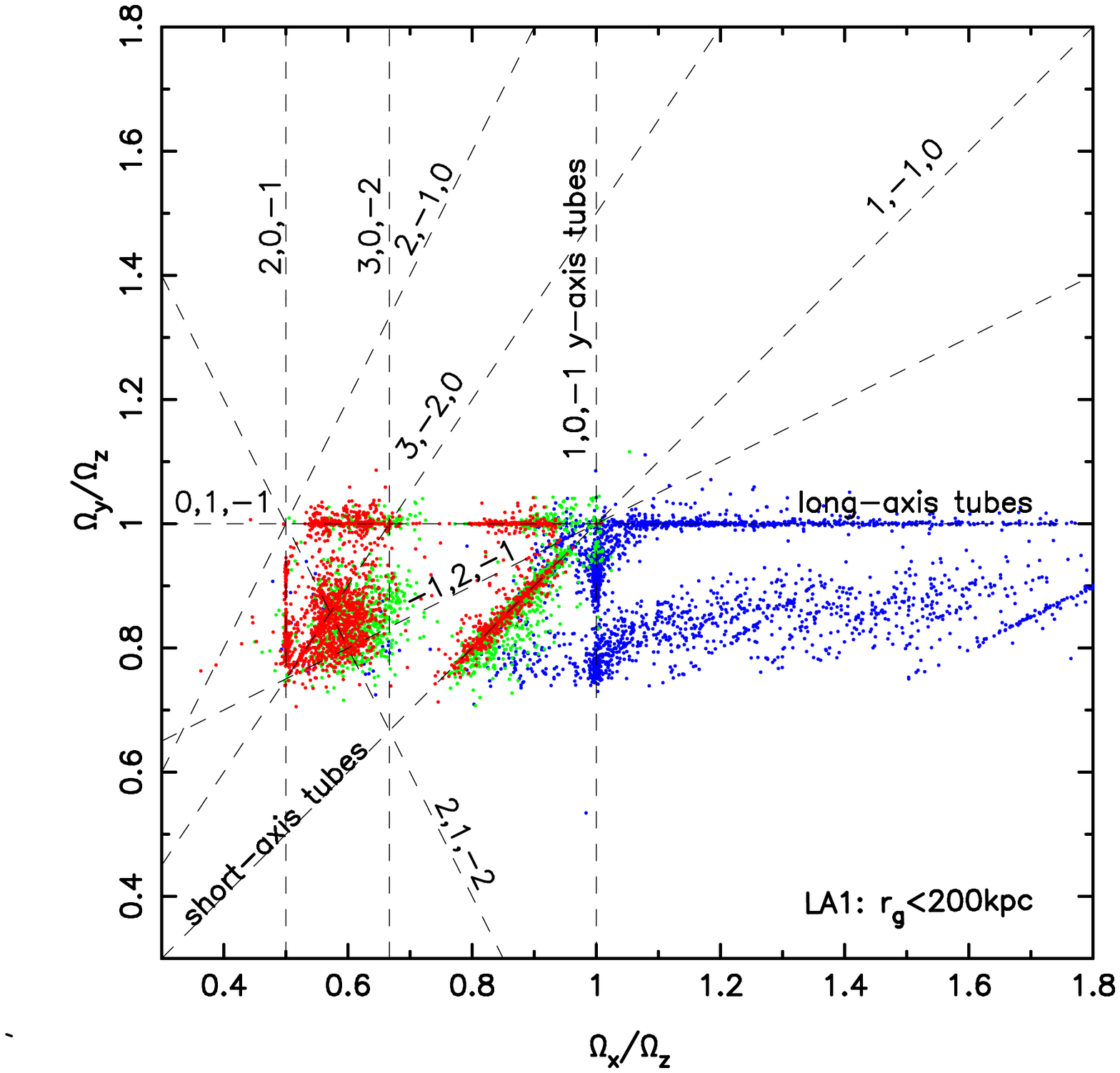}
\includegraphics[trim=0.pt 0.pt 0.pt 0pt,width=0.35\textwidth]{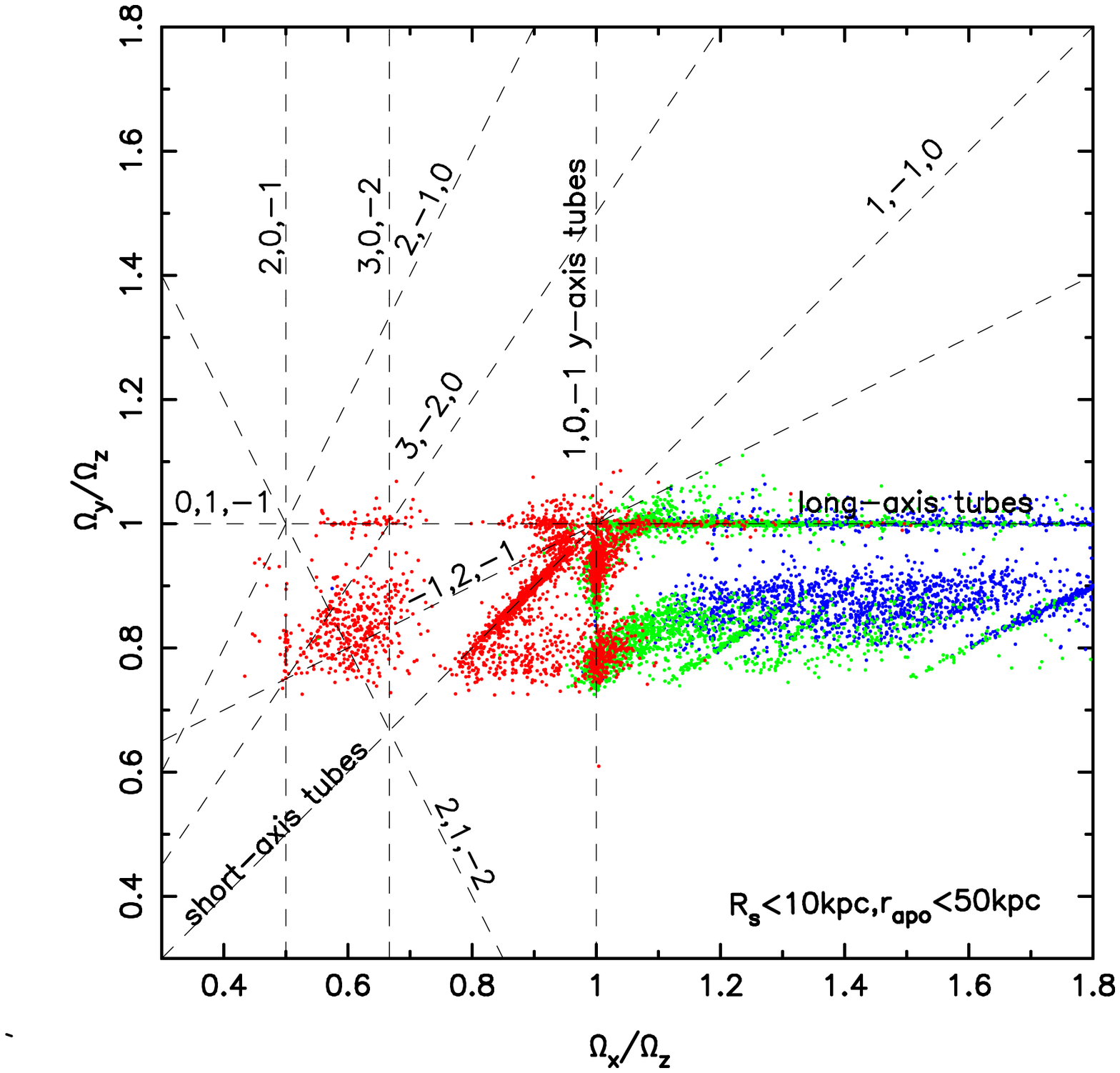}
\caption{Left: Frequency map of $10^4$ halo orbits in LA1 with
  $r_g<200$kpc (the frequency map of these orbits in the original triaxial halo is shown in Fig.~\ref{fig:SA1_map}).  Axes $x,y,z$ chosen to be the {\it global} long,
  intermediate, and short-axes of the halo. Right: frequency map of
  $\sim 10^4$ halo orbits with $R_s<10$~kpc and apocenters less than
  50~kpc. }
\label{fig:LA1_map}
\end{figure*}

 Figure~\ref{fig:SA1_map}~(right) shows a frequency map of $\sim10^4$
 inner halo particles.  A comparison of the right and middle panels of
 Figure~\ref{fig:SA1_map} shows that the main features of the
 frequency map of the entire halo (middle) are also seen in the map of
 inner halo orbits (right).  Thus, although the halo is more oblate
 within 50~kpc that at larger radii, the orbits in the inner halo
 share the major orbit families and resonances of the entire halo. 
 
 The reader may be concerned that since the halo is moderately triaxial and not axisymmetric, by selecting orbits within a ``torus'' of radius $R_s$  rather than in a region localized around the ``sun'',  the frequency maps of stars in the entire torus would not reflect variations in the phase space structure at different locations in the equatorial plane.  We therefore also produced frequency maps of subsets of the full sample of $10^4$
orbits, which contained only those orbits with initial positions within  an individual quadrant of the galaxy model, with quadrants symmetrically about either the major or minor axes. These frequency maps
are not shown since they are virtually indistinguishable from those
of the full samples of $10^4$ orbits (shown in the middle and right
 panels) demonstrating, that within the ``solar'' circle, the halo is sufficiently oblate that the orbital populations do not depend significantly on azimuthal location.

\subsubsection{LA1: Triaxial halo with long axis disk}
\label{sec:LA1}

It has been conjectured that instead of the MW halo being oblate with
short-axis co-aligned with the spin axis of the disk (as in model
SA1), the disk may in fact be prolate and perpendicular to the long
axis of the halo. \citet{helmi_04} concluded, from modeling the tidal
disruption of Sagittarius dwarf satellite, that the kinematics of
stars forming the leading arm of the Sgr stream suggest that the dark matter
halo may be prolate with an average density axis ratio close to 5/3
with long axis perpendicular to the disk. This orientation has also
been suggested by studies of distribution of dark matter subhalos that are
satellites of MW-sized dark matter halos analyzed in cosmological $N$-body
simulations \citep{zen_etal_05}.

We investigated this possibility in simulation LA1, where a disk was
grown perpendicular to the long axis of the halo.  A plot of the
projected density contours of halo particles in simulation LA1, after
the halo relaxed into a new equilibrium with the disk potential are
shown in Figure~\ref{fig:LA1_map}~(left). The density contours show
that the halo triaxiality varies from about 0.8 to 0.4 at small radii
($x <20$kpc), with long axis in the disk plane at these radii but
remains triaxial with long axis perpendicular to the disk at larger
radii.

Figure~\ref{fig:LA1_map} (left) shows a frequency map of $10^4$ halo
orbits randomly selected to lie with $r_g < 200$~kpc (representing the
entire halo DF).  Recall that the frequency map of original triaxial
halo for this model is shown in Figure~\ref{fig:SA1_map}~(left). In
this map a significant fraction of orbits deep inside the potential
(blue) populate the horizontal (1:1) resonance line corresponding to
the global long-axis of the halo. Long axis tube orbits circulate
about the $x$-axis in a fixed direction.  The dramatic increase in the
length and strength of this family of tube orbits as well as their
location on the frequency map (relative to the fraction long-axis tube
orbits in the original triaxial halo in Fig.~\ref{fig:SA1_map} left)
is a direct consequence of the growth of the disk potential. In this
simulation the disk is symmetric about the long $x$ axis. Therefore
orbits (especially those deeper in the potential) experience a
somewhat larger increase $\Omega_x$, than in the other two
frequencies. The greater increase in $\Omega_x$ causes the blue points
on the frequency map to migrate towards the right of the map. Not all
the migration to the right is associated with long-axis tubes, we see a
significant increase in the fraction of box orbits (below the
horizontal line) as well as several distinct resonances in this
population.

We also note that many tightly bound (blue) points clustered along the
vertical line corresponding to the intermediate ($y$) axis tubes. The
``intermediate-axis'' tube family is generally expected to be
unstable. However the strong clustering along the vertical line shows
that this family is both stable and well populated in this
potential. The reason for this is seen in 
Figure~\ref{fig:shape_contours}~(bottom row in 2nd column from left), which shows a
slight elongation of the dark matter (black contours) density along the $y$-axis for $x<10$~kpc. It appears that
in this model the $y$-axis tubes appear for the most bound population
because this is a {\it local long axis}!

In model LA1 there is a large increase in the fraction of tightly
bound (blue) orbits associated with the $x$-axis tube family, which
happens to coincide with the symmetry axis of the disk. This is in
sharp contrast with Figure~\ref{fig:SA1_map} where the disk, oriented
along the $z$-axis caused an increase in the fraction of tightly bound
orbits associated with the ($z$) short-axis-tube family. There are
also short-axis tube orbits in LA1 but most are weakly bound
(red). Thus although the initial triaxial halos were identical in
shape and DF, the differences in the orientation of their disk
relative to the halo resulted in very different orbit populations
(i.e. DFs) especially for the more tightly bound particles.

In Figure~\ref{fig:LA1_map}~(right) we show orbits in the inner halo
($R_s<10$~kpc, $r_{\rm apo}<50$~kpc). The frequency map of the inner
halo (right) is very similar to that of the outer halo (left), with
the main difference being a decrease in the fraction of box
orbits between the resonance lines (2,0,-1) and (3,0,-2). The colors
corresponding to the energies of particles change purely because the
range of energies in the right hand plot is smaller. The box orbit
resonances (below the long-axis tubes and to the right of the $y$-axis
tubes) are seen more clearly, because this region is more densely
populated with orbits due to the orbit selection criterion.

In both SA1 and LA1, the inner halos are significantly flattened in
the plane of the disk (although they remain triaxial) at
$r<50$~kpc. It is therefore remarkable that the frequency maps of
orbits confined to the inner halos of these two potentials (right
panels of Fig.~\ref{fig:SA1_map} and Fig.~\ref{fig:LA1_map}), reflect
the differences in large scale orientation of the halos relative to
the orientations of their disks.  Although there are clearly
differences between the maps of the inner halos and the full halos,
these differences are significantly less than the differences that
arise from the different orientations of the disk. Recall that the
initial halos were identical prior to the growth of the disks. {\it
  This implies that when accurate phase space coordinates for stars
  are obtained from future survey such as Gaia, they can be used to
  infer the global shape of the halo relative to the disk, even if
  accurate coordinates are only obtained within 10~kpc of the sun.}

\subsubsection{TA1: Triaxial halo with tilted disk}
\label{sec:TA1}

\begin{figure*}
\centering
\includegraphics[trim=0.pt 0.pt 0.pt 0pt,width=0.33\textwidth]{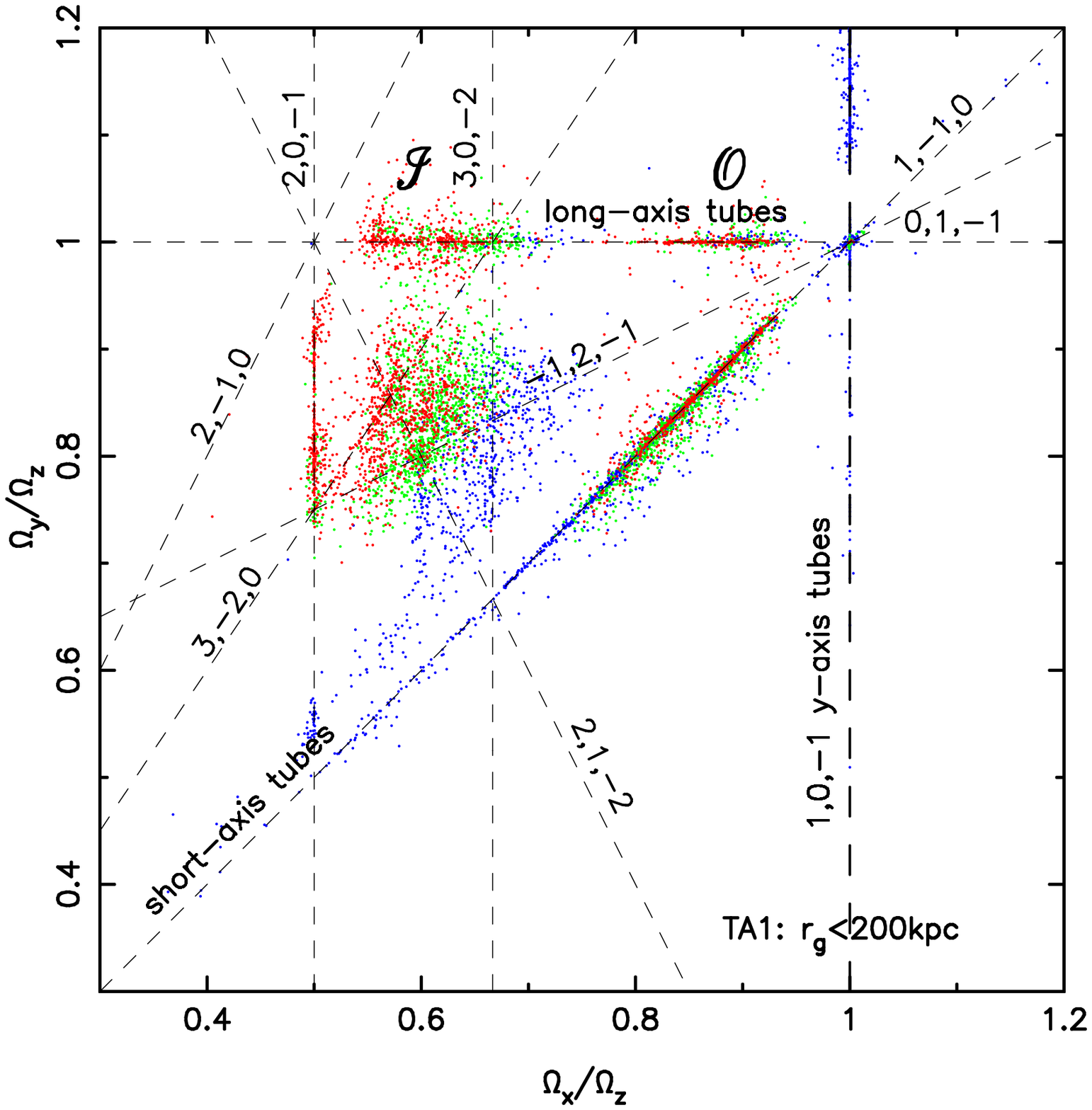}\includegraphics[trim=0.pt 0.pt 0.pt 0pt,width=0.33\textwidth]{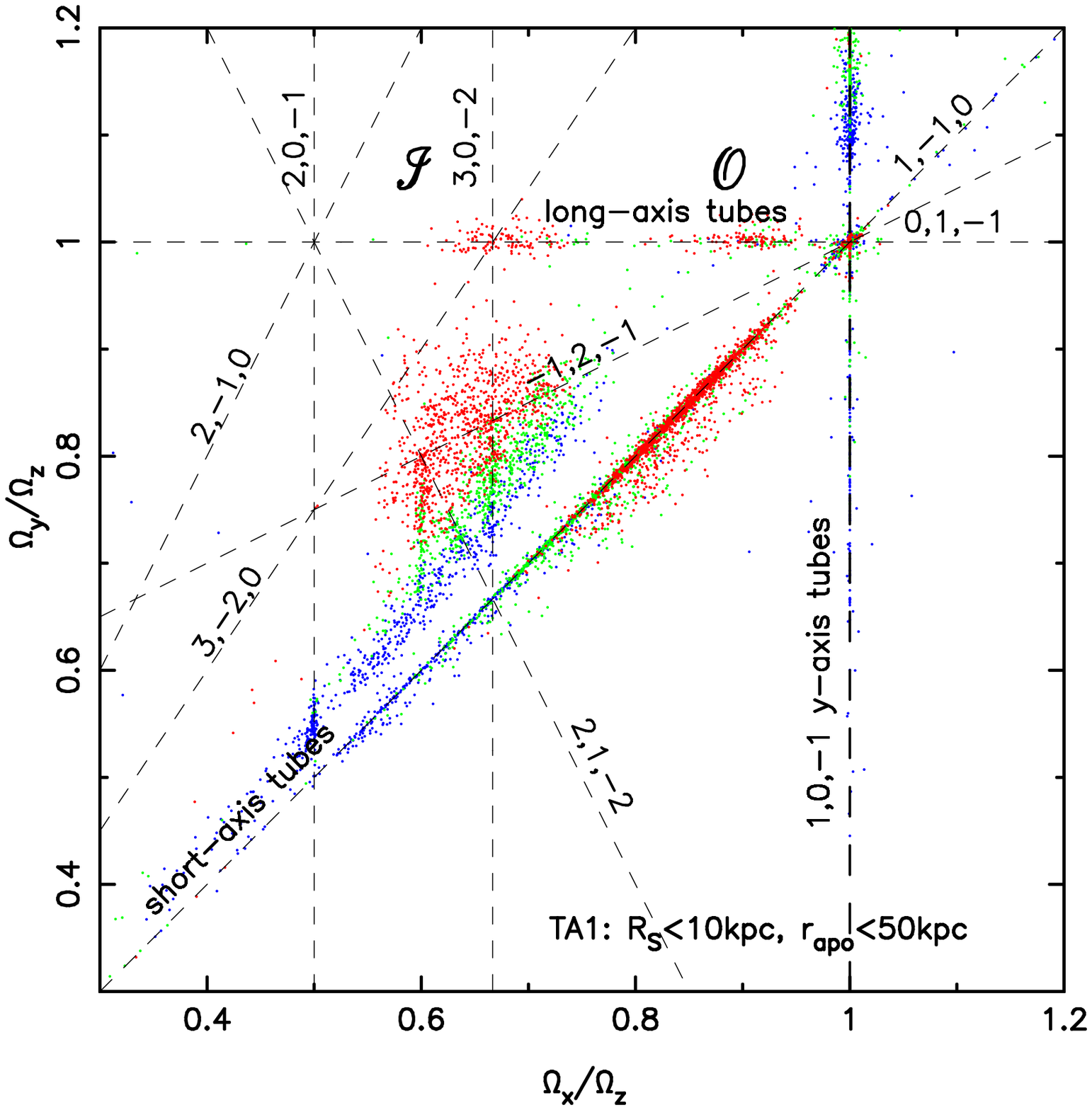}
\caption{{Left}: Frequency map of $10^4$ halo orbits with $r_g
  <200$~kpc after the disk grows tilted at an angle of 30$^\circ$ to
  the $x-y$ plane of the triaxial potential (model TA1) (the frequency map of these orbits in the original triaxial halo is shown in Fig.~\ref{fig:SA1_map} left).  {Right}:
  about $10^4$ orbits selected from the ``solar'' neighborhood and
  confined to the inner halo of the model.}
\label{fig:TA1_map}
\end{figure*}

We now consider a model in which the disk was grown inclined to the
$x-y$ plane of the triaxial halo (model TA1).  Such an orientation is
motivated by studies of dark matter halos from cosmological $N$-body
simulations that show that the relative orientation of the angular
momentum axis of triaxial halos is on average no more than
$25^\circ-30^\circ$ from the short axis of the triaxial halo
\citep{bailin_steinmetz_05}. Furthermore, numerous simulations show that the
disk can be misaligned with the symmetry axis of the halo
\citep{vandenbosch_etal_02}.

Figure~\ref{fig:TA1_map}~(left) shows the frequency map of $10^4$ halo
particles with $r_g <200$~kpc. This frequency map is very similar to
the model with the short-axis disk (Fig.~\ref{fig:SA1_map} (middle)),
but also shows a clustering of (blue) points along a vertical line at
$\Omega_x/\Omega_z =1$ (corresponding to $y$-axis tubes). This family
of orbits arises because the disk axis in this model is inclined to
the $x-y$ plane such that the $x$-axis is in the plane of the disk but
the $y$-axis is at an angle of 30$^\circ$ to the disk. The disk
potential therefore induces resonant trapping of tightly bound halo
orbits causing a larger circulation in the plane of the disk, and
consequently more angular momentum about the $y$-axis.  Although this
family is rotating about an axis inclined to the $y$ axis, this
trapping appears on the Cartesian frequency map as an enhanced
clustering of tightly bound orbits about this axis. This family of
intermediate axis tubes does not appear in the most tightly bound
orbits (blue) of any other model. Unlike in model LA1, the density
contour plots do not show a noticeable elongation along the $y$-axis.

The DF of the inner halo of TA1 (Fig.~\ref{fig:TA1_map}~right)
displays many of the features of the full DF within 200~kpc (middle),
with some differences: the inner long-axis tube family is
less prominent, the banana family (2,0, -1) is very sparsely populated
in the inner region and a new resonance family is seen at
$\Omega_x/\Omega_z=0.6$. Apart from these  differences, the
important major orbit families (the long-axis tubes, short-axis tubes
and intermediate-axis tubes) are well represented, showing that the DF
of the inner halo, while different from that of the entire halo,
shares the most important characteristics that distinguish it from
other models.

Figure~\ref{fig:shape_param} shows that the shape of the inner halo of
TA1 (grey curves) is very similar within 50~kpc, to the shape of the
inner halo of SA1 (black curves).  However the frequency map of the
inner halo of TA1 (Fig.~\ref{fig:TA1_map}~right) is quite different
from that of SA1 (Fig.~\ref{fig:SA1_map}~right). This implies that
their DFs are very different - again a consequence of the different
original orientation of the disks relative to their globally triaxial
halos. Thus, our analysis of the orbital phase space structure using
the frequency maps allows us to gain insights into the differences in
the orbital structures of two halos with very similar inner halo
shapes.

\subsubsection{IA1: Triaxial halo with intermediate axis disk}
\label{sec:IA1}
\begin{figure*}
\centering
\includegraphics[trim=0.pt 0.pt 0.pt 0pt,width=0.33\textwidth]{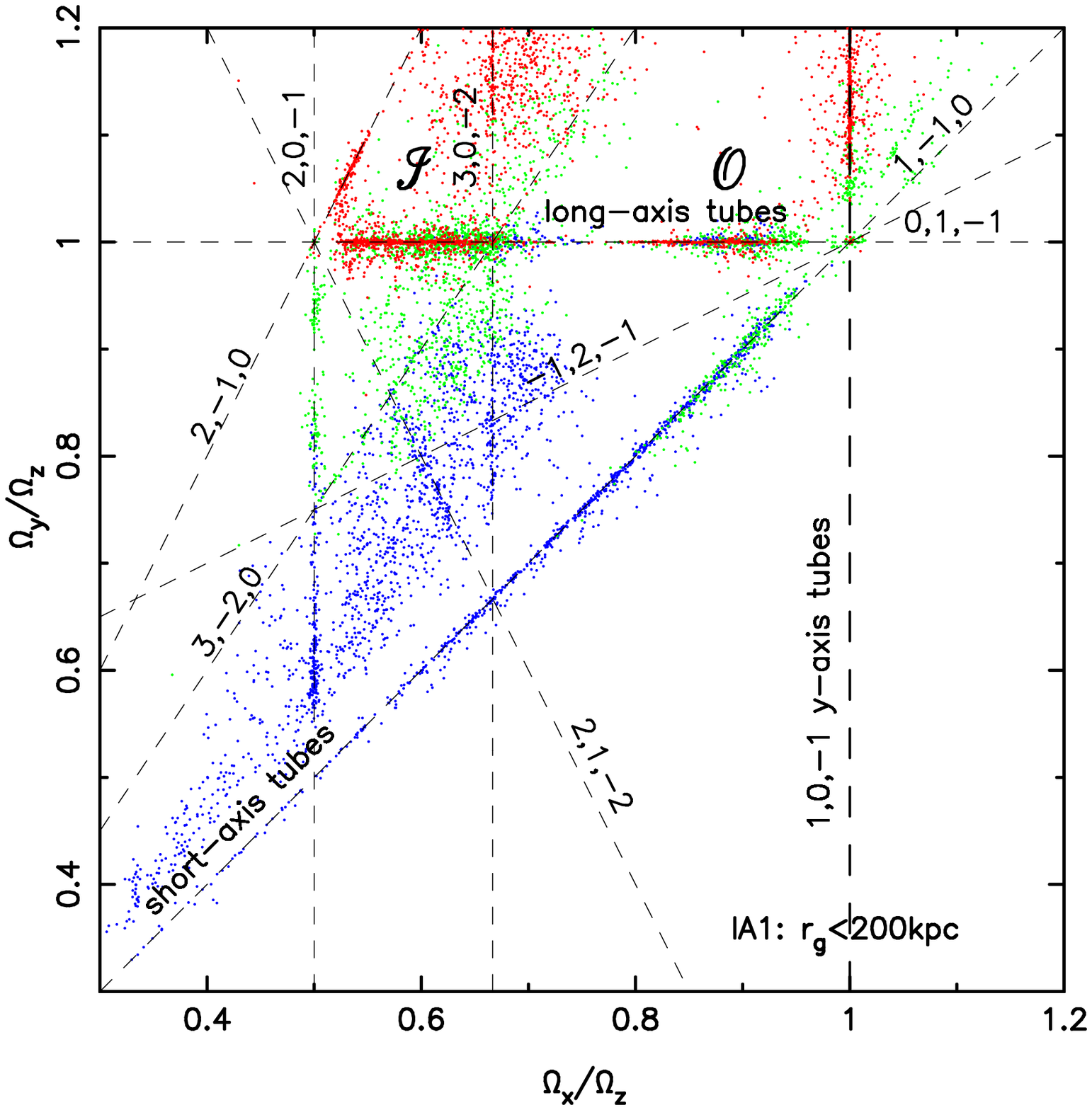}
\includegraphics[trim=0.pt 0.pt 0.pt 0pt,width=0.33\textwidth]{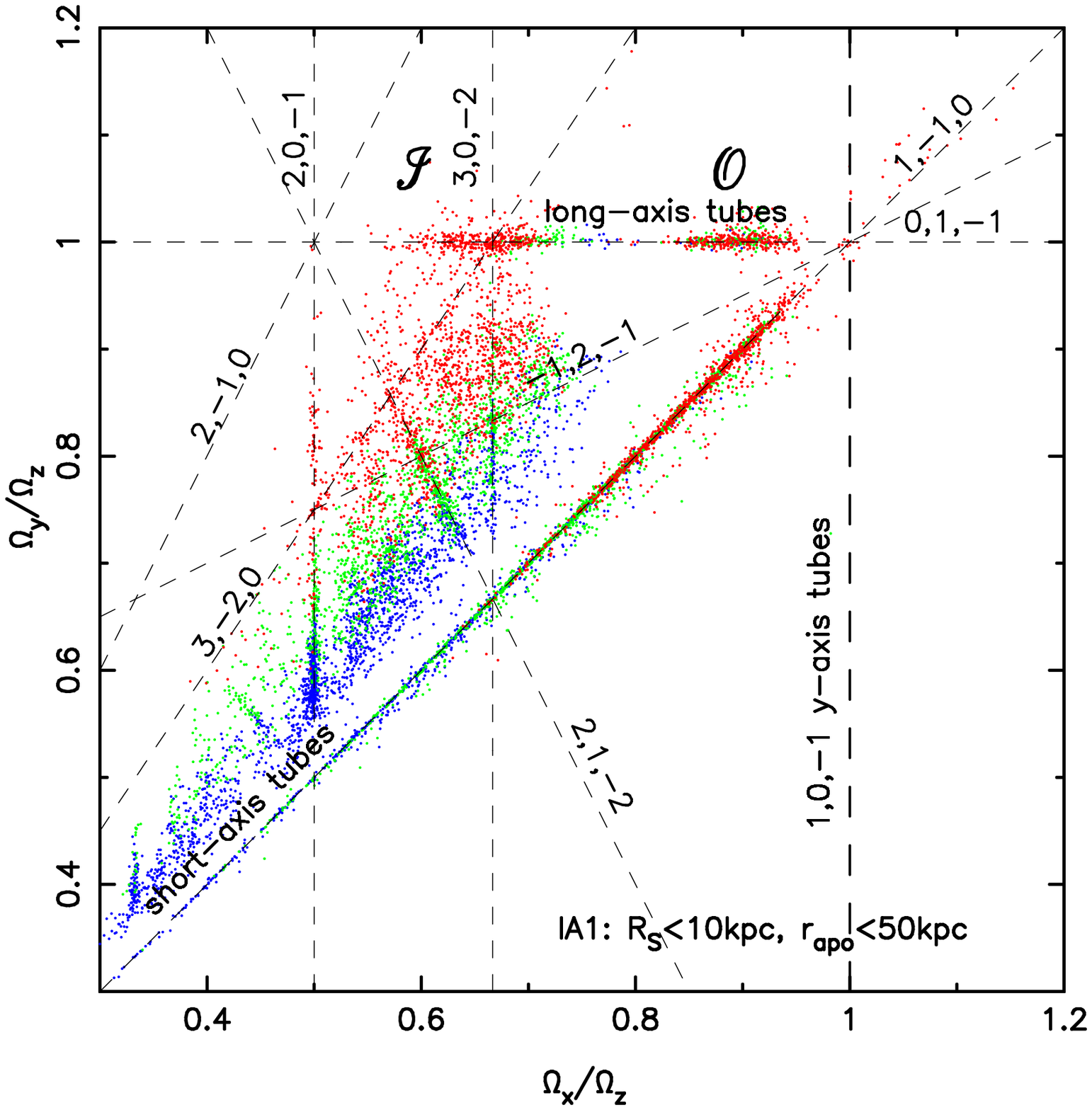}
\caption{{Left}: Frequency map of $10^4$ halo orbits with $r_g
<200$~kpc after the disk grows perpendicular to the intermediate ($y$)
axis in the triaxial potential (model IA1).  The frequency map of these orbits in the original triaxial halo is shown in Figure~\ref{fig:SA1_map} (left). {Right}: about $10^4$
orbits from the inner halo of the same model. The $y$-axis tube family
is populated by weakly bound (red) points and is not important in the
solar neighborhood map on the right.}
\label{fig:IA1_map}
\end{figure*}

Recently, Law and collaborators \citep{law_etal_09, law_majewski_10}
remeasured the shape of the MW halo by fitting both the velocities and
positions of stars in the Sagittarius tidal stream with a triaxial
halo model. They find that a slightly triaxial halo with the sun
located roughly along the minor axis gives the best fit to the
available kinematic and positional data of stream stars. The best-fit
configuration requires the MW disk to be perpendicular to the
intermediate axis of the triaxial halo - a disk configuration that is
believed to be inherently unstable \citep{heiligman_schwarzschild_79}.

Our final model simulates this configuration with a rigid disk
potential grown perpendicular to the intermediate axis of the triaxial
halo (models investigating the stability of halos with live disks will
be presented in \citet{deb_etal_11}).  The frequency map of $10^4$
randomly selected halo orbits with $r_g<200$~kpc is shown in
Figure~\ref{fig:IA1_map}~(left). The map shows a prominent clustering
of weakly bound orbits along the vertical line corresponding to
intermediate axis tubes with $\Omega_x/\Omega_z=1$ at values of
$\Omega_y/\Omega_z>1$. Since the disk is symmetric about the $y$-axis,
at small radii the halo does become more oblate, with $y$ as its
symmetry axis. However we see that even the weakly bound particles
(red), which are at large radii and expected to be less affected by
the disk, are also associated with the intermediate ($y$)-axis tube
family. Figure~\ref{fig:shape_param} shows that although the shape of
halo SA1 (black curves) is more oblate than halo IA1 (red curves), the
radial profiles (of $b/a$ and $c/a$) are very similar for these two
halos (except in the absolute degree of flattening). Consequently,
when we plot only those orbits which are confined to the inner halo
(Fig.~\ref{fig:IA1_map}~right), we see that the frequency maps of IA1
and SA1 (Fig.~\ref{fig:SA1_map}~right) are so similar that they are
hard to distinguish from each other. The intermediate axis tube family
that was seen in Figure~\ref{fig:IA1_map}~(left) has completely
disappeared, showing that all the orbits that made up this family were
part of the outer halo. However, this similarity in the distribution
functions is not entirely surprising since
Figure~\ref{fig:shape_contours} show that SA1 and IA1 have similar
density contour distributions (this similarity was also found by D08).

\subsubsection{Discussion of disk-halo orientation effects}
\label{sec:orient}

Figures~\ref{fig:shape_param} and \ref{fig:shape_contours} showed that
the growth of a disk galaxy in a triaxial dark matter halo of
arbitrary orientation modifies the shape of the inner part of halo,
but leaves the outer part largely unaffected. However most methods for
determining the shape of the halo, assume that dark matter is
stratified on concentric similar ellipsoids of constant shape
(i.e. not varying with radius). With this assumption the shape of the
halo can be measured with an accuracy of a few percent, out to 50 -
70~kpc \citep{johnston_etal_96, gnedin_etal_05}. Since the shape of
the halo probably varies significantly with radius due to disk
formation, the assumption of a constant halo shape is not valid.

Although the radial variation of the shape of the halo will be
impossible to measure, the fact that in all cases the inner halo is
nearly oblate and flattened like the disk will enable us to use halo
orbits to constrain both the shape and the DF of the inner halo.

Furthermore, the analysis of the DFs of four different halo models
showed that a frequency map provides detailed information on the
various orbit families that constitute the halo DF and relative
abundance of the different orbit families at various energies. In
addition, while the disk potential traps tube orbits, which share the
disk symmetry axis. Even restricted maps of halo orbits confined to
the inner region ($r_{\rm apo} \sim 50$~kpc) show all the orbit
families present in the global halo.

It is worth noting that frequency maps are a superior method for
identifying important orbit families in a potential than the standard
methods that rely on the properties of specific orbit families
\citep{carpintero_aguilar_98,valluri_etal_10, deibel_etal_11}. This is
because the standard methods of orbit classification, specify the set
of orbital types that will be used to classify orbits. They assume
{\it a priori} that the shape of the halo is constant with radius. In
triaxial potentials with constant shape, intermediate axis tubes are
unstable and not expected in the DF
\citep{heiligman_schwarzschild_79}, yet we see from the frequency maps
that three of four models contained members of this family, because
the triaxiality of the halo varies with radius. To build a DF for the
stellar halo, it is important to have a full representation of all the
orbit types that are important in the halo, since the orbit
populations, in turn, reflect the large scale shape, orientation and
formation history of the Galactic halo.

We caution that in all the simulations above, the disk particles were
held fixed and did not dynamically respond to the change in the halo.
Theoretical arguments indicate that at least some of these
orientations, e.g.  IA1 \citep{heiligman_schwarzschild_79}, are likely
to be dynamically unstable. A more detailed analysis of
self-consistent dynamical models of such systems are needed to
ascertain whether such disk orientations would be found in nature
\citep{deb_etal_11}.

\subsection{Non-axisymmetric halos with live disks}
\label{sec:live_disk}

\begin{figure*}
\centering \includegraphics[trim=0.pt 0.pt 0.pt
  0.pt, width=0.33\textwidth]{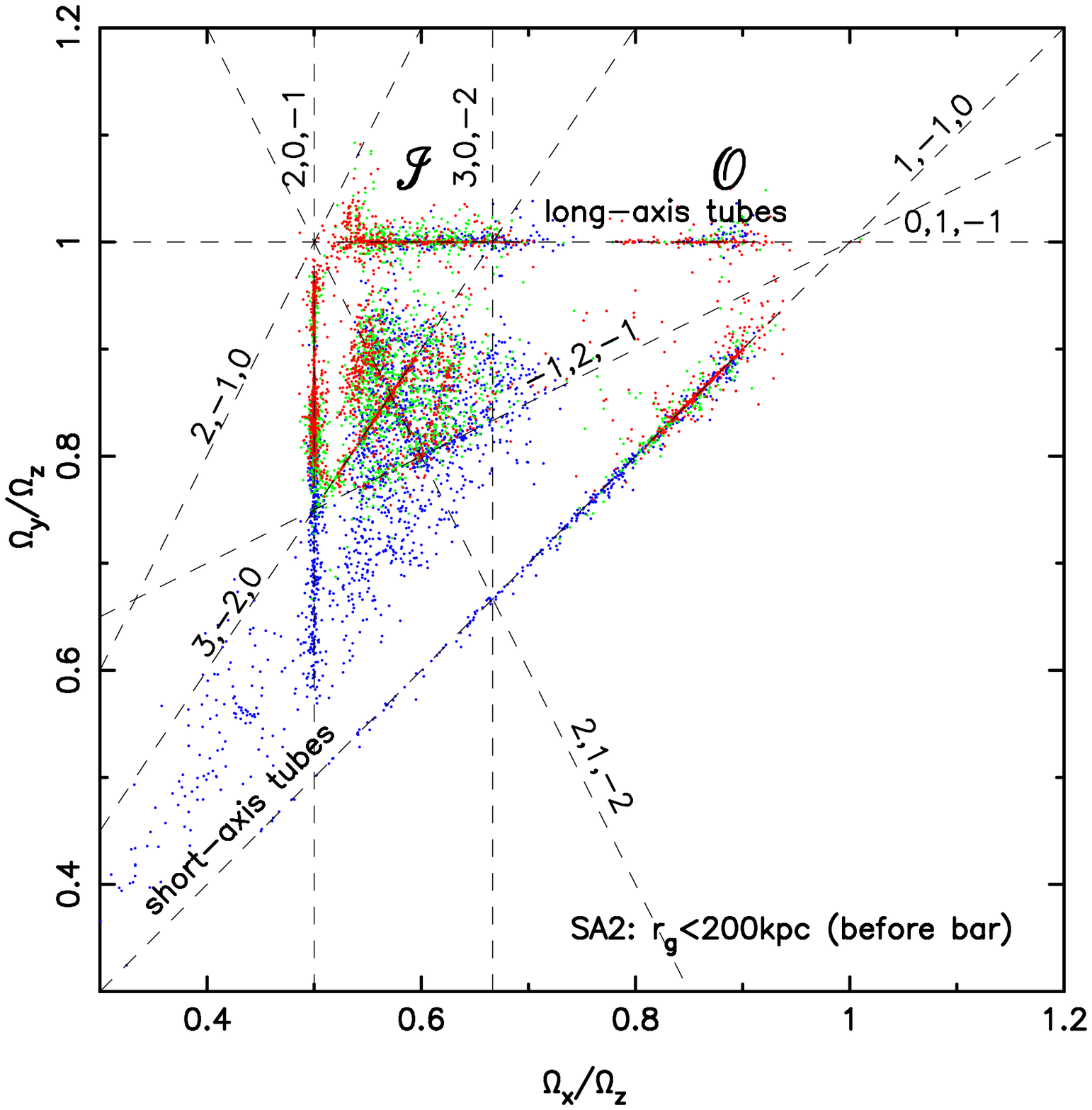}
\includegraphics[trim=0.pt 0.pt 0.pt
  0pt,width=0.33\textwidth]{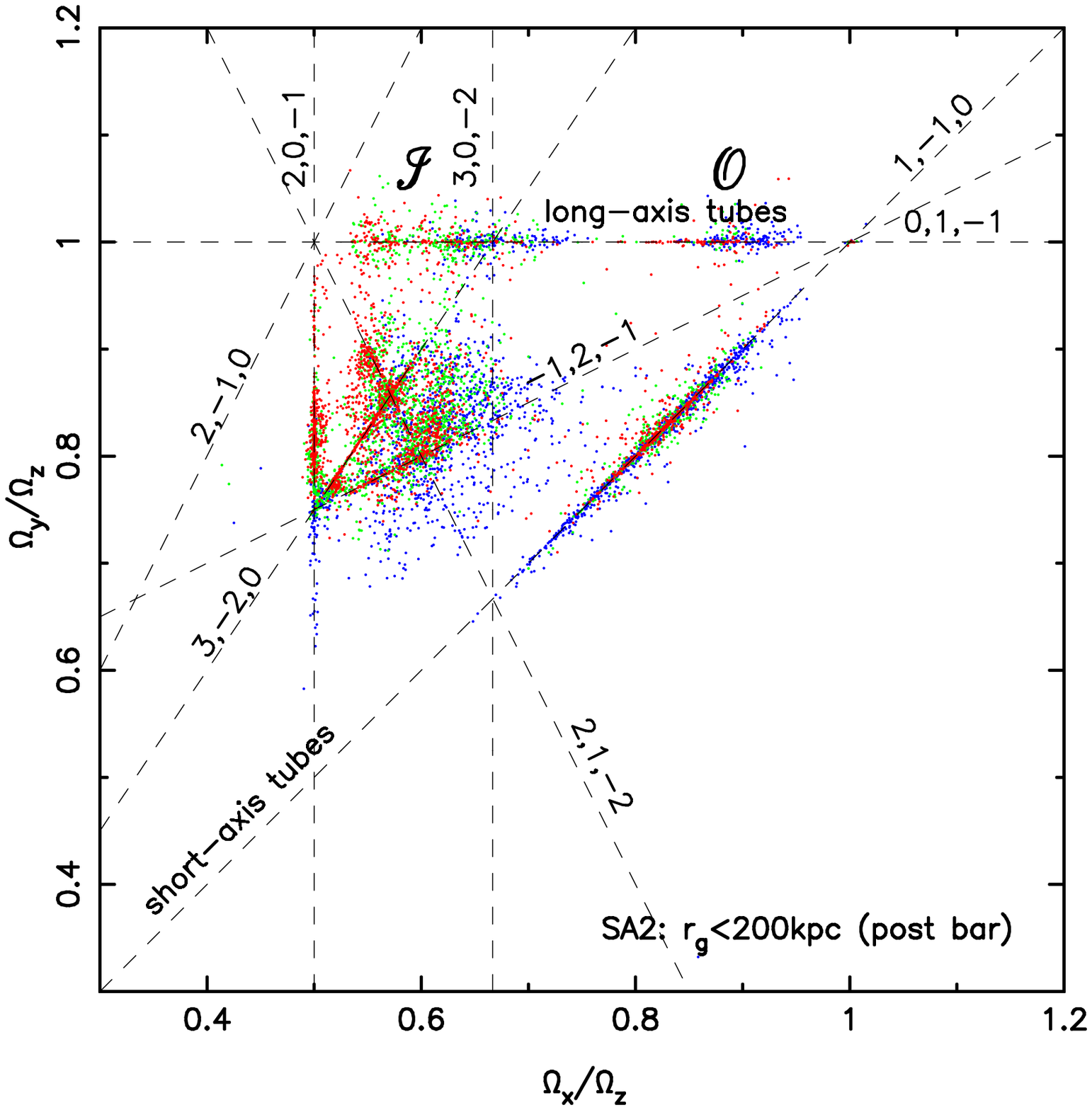}
  \includegraphics[trim=0.pt 0.pt 0.pt
  0pt,width=0.33\textwidth]{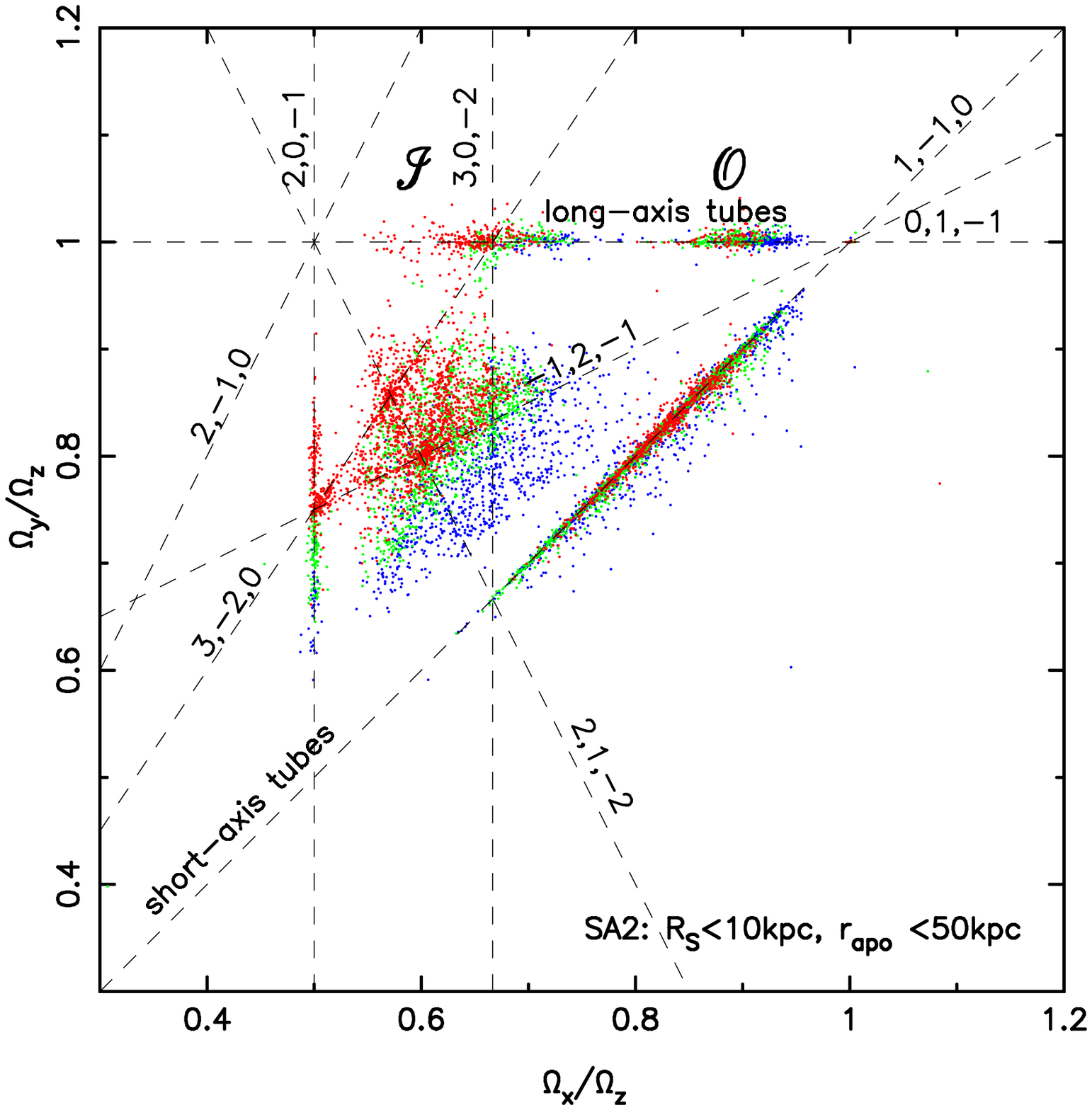}
\caption{Left: Frequency map of $10^4$ halo orbits with $r_g<200$~kpc
  after a short axis disk is grown in the triaxial halo (model SA2)
  before the formation of a bar. The frequency map of these orbits in
  the original triaxial halo is shown in Fig.~\ref{fig:SA1_map}
  (left). The disk of live particles rapidly forms a bar, which
  subsequently dissolves. Middle: frequency map of the same halo
  orbits after the bar has dissolved. Right: frequency map of $10^4$
  orbits in the inner halo after the bar has dissolved.}
\label{fig:SA2d_map}
\end{figure*}

Most stellar disks, including that of the MW, are not axisymmetric or
time-independent since they contain features such as spirals and
bars. These non-axisymmetric features drive angular momentum
exchanges.  The angular momentum exchange that occurs between a
rotating bar and the dark matter (and presumably also the stellar
halo) is expected to result in a change in the angular momentum of
halo particles that interact with the bar \citep{weinberg_85,
  deb_sel_00}. In this section we study the effects of angular
momentum and energy transfer resulting from coherent time-dependent
perturbations from a bar. We also studied two models in which a
stellar and gaseous disk forms via realistic processes in a spherical
halo and in a prolate halo. In these simulations, hot gas in the
initial halo which has angular momentum, cools to form a rotating disk
followed by star formation and feedback (see \S~2 for details).

\subsubsection{Triaxial halo with barred disk}

To study the effects of a bar, we adiabatically grew a stellar disk
that had only $30\%$ of the mass of the disk in the previous
simulations inside a triaxial halo (with symmetry axis parallel to the
short axis of the halo, model SA2).  Because of its low mass, the bar
that forms in this system is eventually destroyed by its interaction
with the triaxial halo
\citep{berentzen_shlosman_06}. Figure~\ref{fig:SA2d_map}~(left), shows
the frequency map of $10^4$ halo particles with $r_g < 200$~kpc after
the disk had reached full mass. The frequency map shows the two
long-axis tube families (marked with script ``$\mathscr I$'' and
``$\mathscr O$''), a strong short-axis tube family and several
box-orbit resonances (e.g. (2,0,-1), (2,1,-2), (3,-2,0), as well as a
few other unlabeled resonances seen mostly in red points).

After the disk reached its full mass the particles were made ``live'',
i.e. the disk was allowed to evolve self-consistently along with the
halo. The rotating disk rapidly formed a bar and spiral patterns. The
bar survived for $\sim$10~Gyr and finally dissolved. Note that we
always integrate orbits in a potential without a bar.  This is because
we ``freeze'' the potential before integrating orbits, and the
presence of a bar would result in an unrealistic ``freezing'' of a
time-dependent rotating bar. Most models that seek to obtain the DF of
the Milky Way disk and halo \citep[e.g.][]{binney_10,
binney_mcmillan_11}, neglect the bar since it is difficult to model
its time-dependent potential.

Figure~\ref{fig:SA2d_map}~(middle), shows a frequency map of the same
set of $10^4$ halo particles with $r_g<200$~kpc, while the right panel
shows $10^4$ halo particles in the inner halo, after the growth and
dissolution of the bar. This frequency map shows that tightly bound
(blue) particles that were associated with the (2,0,-1) resonance and
the short-axis tube family are scattered to other parts of the
map. This is because these orbits are most strongly affected by the
central bar. The bar exchanges angular momentum with resonant halo
particles \citep{tremaine_weinberg_84_dynfric,deb_sel_00,
  athanassoula_02}.  Thus the blue (tightly bound) halo particles in
the bottom right of the left panel are scattered and the resonances at
the bottom left of the map are no longer seen.  Many of the blue
points now appear to be associated with the outer long-axis tube
family (at $\Omega_x/\Omega_z \sim 0.9$) and the short-axis tube
family.  Resonances associated with less bound (red) particles have
become broader or disappeared. Other resonance lines populated by the
weakly bound (red) particles are only slightly affected (e.g. 2,0,-1
and 3,-2,0). These tightly bound particles are seen more clearly in
the right panel which shows orbits from the inner halo.

 It is clear that although the middle and right plots have fewer
 strong resonances than the halo prior to the bar (left), the major
 orbit families still appear with the same relative strengths.
 Although some resonances appear broader and those associated with
 particles deeper in the potential have been scattered to other
 locations on the map, the overall structure of the frequency maps are
 similar to SA1 which is also a model with a disk rotating about the
 short-axis of triaxial halo. This shows that although time dependent
 perturbations from a bar can affect the shape and DF of the halo,
 many resonances survive, especially those that are populated by
 orbits that do not interact strongly with the bar, and the overall
 structure of the frequency map is characterized  by the global nature of
 the potential.

\subsubsection{Halos with live disks of gas and stars}
\begin{figure}
\centering \includegraphics[trim=0.pt 0.pt 0.pt 0pt,width=0.34\textwidth]{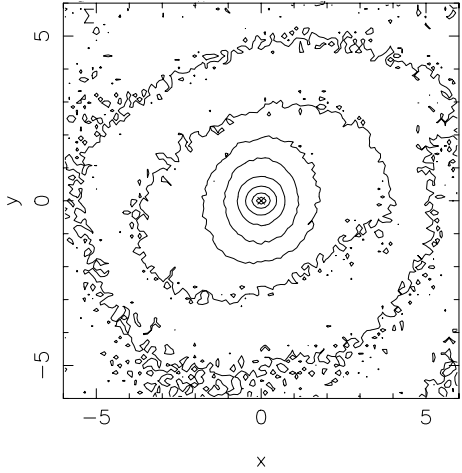}
\caption{Density contours of disk that forms in the spherical halo from cooling halo gas. The disk develops a mild oval distortion ($x$ and $y$ axes coordinates are in kpc).}
\label{fig:contour_SNFWgs}
\end{figure}

We now study frequency maps of halo orbits from models where a stellar
disk forms from hot gas with some initial angular momentum (defined to
be about the $z$ axis) in a spherical halo (model SNFWgs) and in a
prolate halo (model SBgs) (see \S~\ref{sec:simulations} for details).
Following the formation of the disk, the dark matter halo became
fairly oblate in model SNFWgs, while the disk develops only a mild
oval distortion (see Fig~\ref{fig:contour_SNFWgs}).  In these
simulations no stars form in the halo and therefore we once again
study the orbits of dark matter particles.

\begin{figure*}
\centering
 \includegraphics[trim=0.pt 0.pt 0.pt 0pt,width=0.42\textwidth]{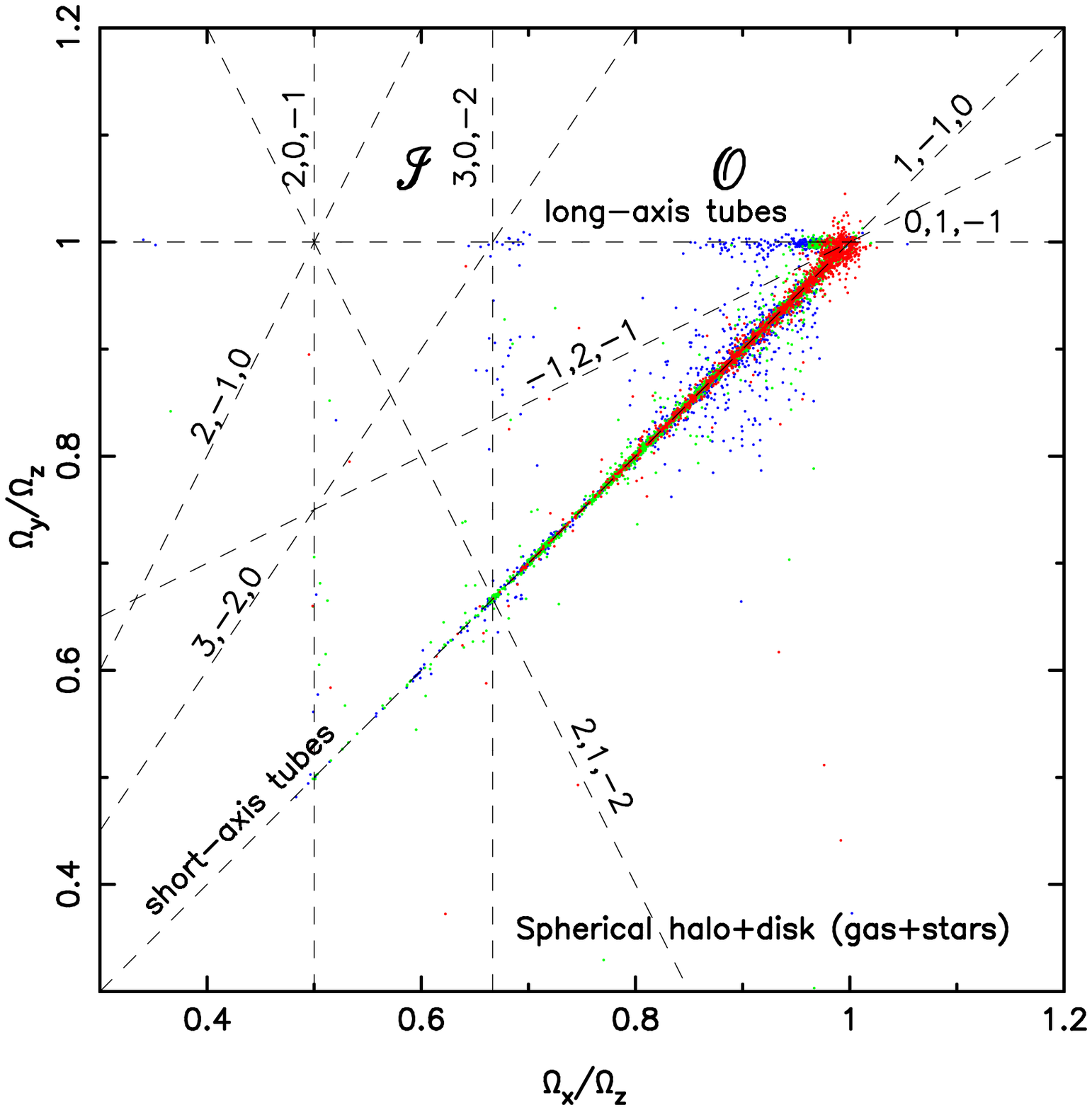}
\includegraphics[trim=0.pt 0.pt 0.pt  0pt,width=0.42\textwidth]{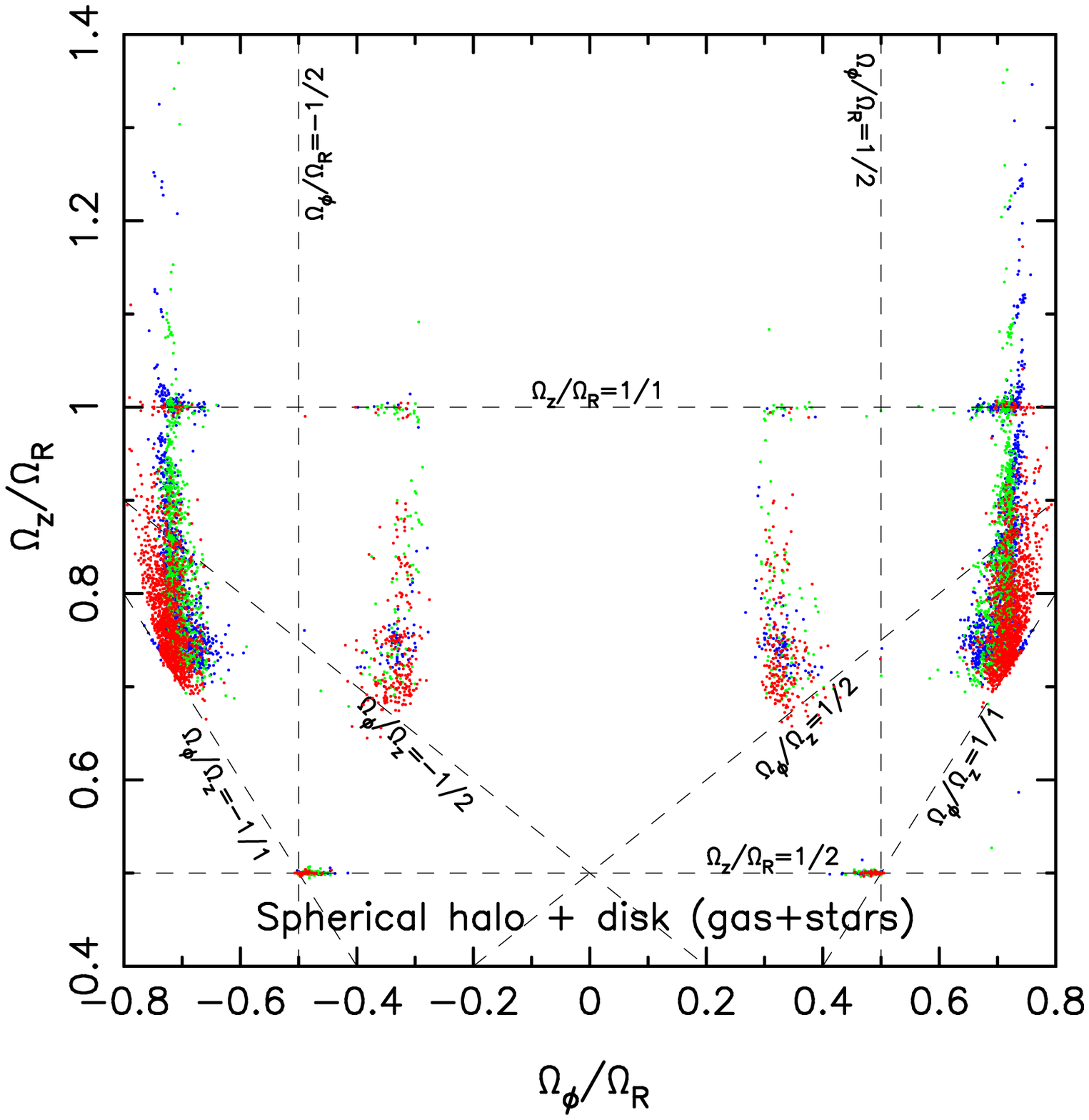}\\
 \includegraphics[trim=0.pt 0.pt 0.pt 0pt,width=0.42\textwidth]{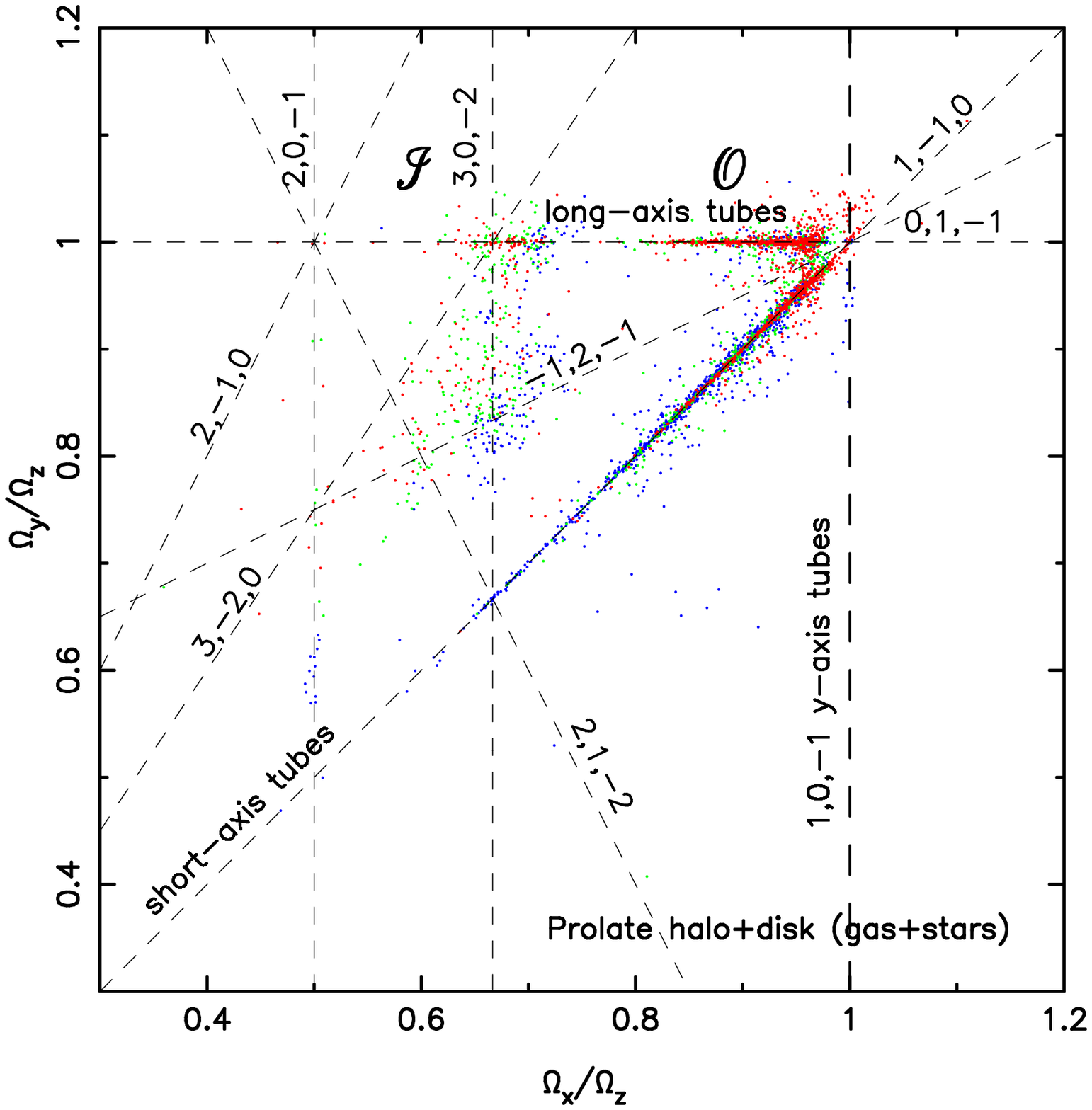}
\includegraphics[trim=0.pt 0.pt 0.pt
  0pt,width=0.42\textwidth]{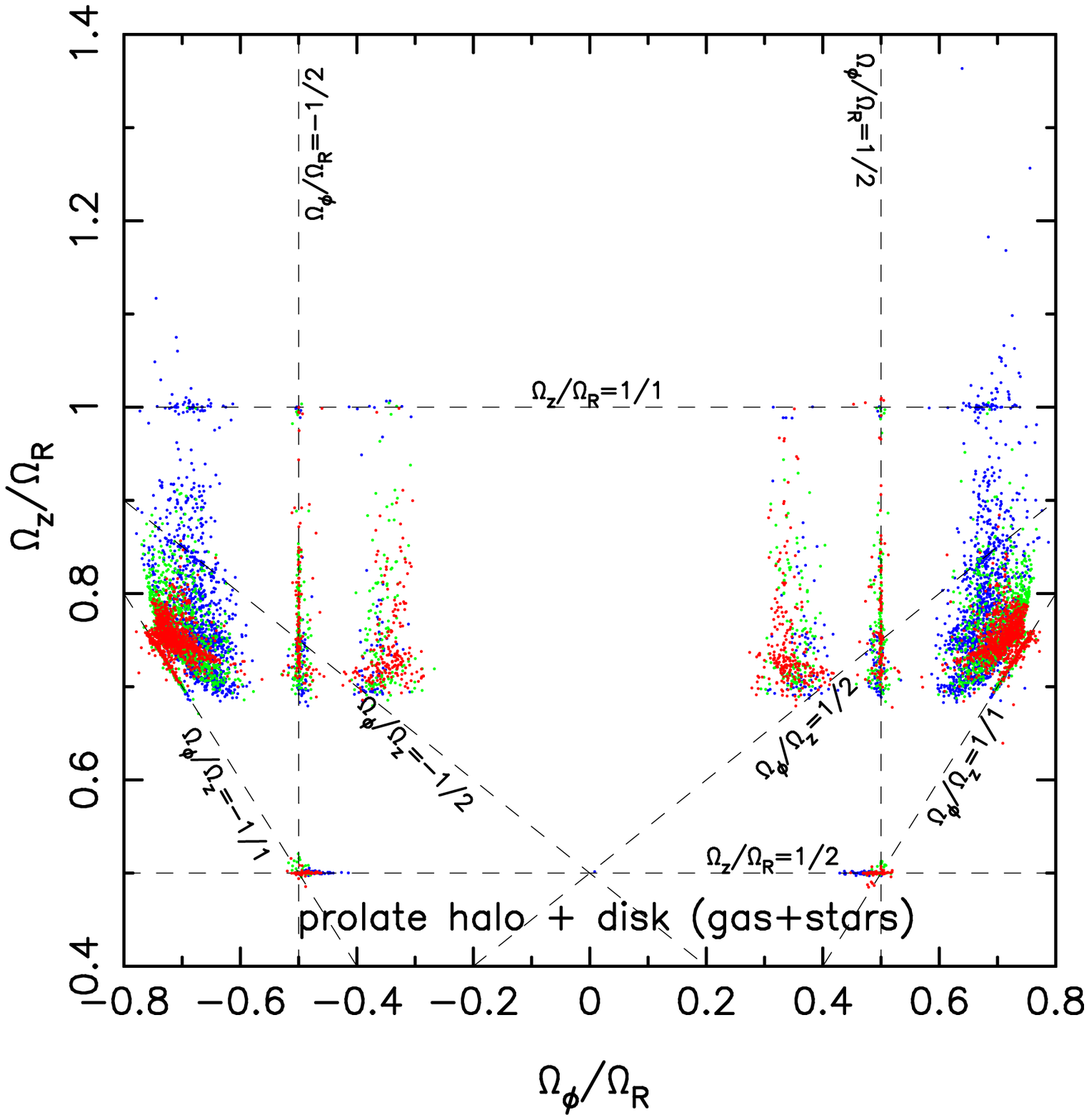}
\caption{Frequency maps of $10^4$ halo particles with $r_g <200$~kpc
  in two halos in which hot halo gas cools into a star-forming disk.
  Left: frequency maps of halo orbits in Cartesian coordinates; Right:
  frequency maps of the same particles in cylindrical coordinates. Top
  row: maps of halo orbits in the case when the initial halo was
  spherical NFW, the bottom row shows frequency maps in the case when
  the initial halo was prolate axisymmetric.  Both Cartesian maps show
  short-axis tubes and long-axis tubes, and only a few box
  orbits. Cylindrical maps show several resonances as indicated by
  thin dashed lines and labels.}
\label{fig:661_gasdk}
\end{figure*}

Figure~\ref{fig:661_gasdk}~(left panels) shows the frequency maps in
Cartesian coordinates of $10^4$ halo stars selected with $r_s<200$~kpc
in the initial spherical halo (top) and in the initial prolate halo
(bottom). As we saw in several previous models (e.g. SA1, TA1), the
growth of the disk results in an increase in the fraction of
short-axis tubes deep in the potential (blue points along the
diagonal). Both models show prominent short-axis tube families which
manifest as the strong clustering along the diagonal.

V10 showed that prolate halos are dominated by long-axis tubes which
persist even after the growth of a baryonic component, but these
long-axis tube orbits become ``rounder''. Long-axis tubes are seen in
the prolate halo as the clustering of points along the horizontal line
with $\Omega_y/\Omega_z =1$ (bottom left panel). Surprisingly the
spherical halo also shows long-axis tubes (predominantly in the
tightly bound orbits colored blue). This is likely to be because the
disk forms a slight oval distortion (see Fig.~\ref{fig:contour_SNFWgs}).  
Compared to the originally triaxial
models SA1, LA1, IA1 and TA1, these two models have only a small
fraction of box orbits (scattered points in the middle of the map)
because these models are not triaxial.

It is also instructive to analyze the same set of halo orbits in
cylindrical coordinates (with $z$ as the symmetry axis)\footnote{This
  coordinate system is not the proper one in which to examine
  long-axis tubes, which require cylindrical coordinates where $x$ is
  the axis of symmetry.}.  The two right panels of
Figure~\ref{fig:661_gasdk} show frequency maps in cylindrical
coordinates. Several resonances previously seen in
Figure~\ref{fig:SphNFW_map}, such as $\Omega_z/\Omega_R = 0.5, 1$, are
seen in both the spherical halo and the prolate halo. In addition both
show new resonances $\Omega_\phi/\Omega_z = \pm 0.5, 1$.  The prolate
halo (bottom right) also shows a large fraction of particles
associated with the vertical resonance line at $\Omega_\phi/\Omega_R =
\pm 0.5$, which are long-axis tubes.

The disk formation process in this subsection is dynamically quite
different from that in the previous section where rigid disks made up
of particles were adiabatically grown in place. These two simulations
confirm that resonant trapping of halo orbits occurs even during the
more realistic dissipative processes by which real stellar disks form.

\section{Using frequency analysis to constrain potential parameters} 
\label{sec:potential}

In the simulations so far, we have integrated orbits in the total
$N$-body potential for the galaxy model, which was known
perfectly. Since the orbits were integrated in the self-consistent
$N$-body potential from which they were drawn, the frequency maps
obtained represent the true DFs of these galaxy models.  Since a major
goal of current and future galactic surveys is to obtain both the
potential and DF of the Galaxy \citep{binney_10,binney_mcmillan_11},
our frequency based method is a promising approach for doing
this. However, in reality the potential of the Galaxy is not known and
it will also be measured from the spatial and kinematic distribution
of stars. Ideally both the self-consistent DF and the potential will
be recovered from six phase space coordinates for large numbers of
stars \citep{binney_10, binney_mcmillan_11}.

The most promising methods for measuring the potential of the halo use
kinematics of stars in tidal streams \citep{johnston_etal_99,
  mcmillan_binney_08}, or accurate orbits of hyper-velocity stars
\citep{gnedin_etal_05}.  While these methods are expected to yield
excellent estimates of the shape and density profile of the MW halo if
it is stratified on concentric ellipsoids, all current numerical
experiments show that the halo's shape varies with radius, making the
measurement of the shape of the halo at all radii challenging due to
the absence of coherent tidal streams and or hypervelocity stars over
a range of radii.

Halo stars are however, distributed over a wide range of radii and
current studies show that even local halo stars have enough kinetic
energy to travel to large radii \citep{carollo_etal_10}.  We now test
whether it will be possible to gain information about the true
potential and DF from the frequency analysis of a large number of such
halo stars.  Jeans Theorem states that any steady state
equilibrium distribution function depends on phase space coordinates
only though the integrals of motion \citep{BT}. This implies that for
a DF (or a random subsample thereof) that is in self-consistent
equilibrium with its background potential, only a small fraction of
orbits are irregular or chaotic.  If a large fraction of orbits are
chaotic (i.e. not confined to regular tori in phase space), they are
expected to diffuse in phase space causing the DF to evolve until the
system reaches a new equilibrium \citep{merritt_valluri_96}. If the
degree of chaoticity is low (either there are only a small number of
strongly chaotic orbits, or there are many weakly chaotic orbits) the
evolution of the potential could take some time and fairly long-lived
quasi-equilibria can exist \citep{poon_merritt_04}. However, if the
initial conditions of a large number of orbits are strongly out of
equilibrium with the background potential, the overall degree of
chaoticity (or ``chaotic momentum'') is large and the self-consistent
system will evolve rapidly \citep{kalapotharakos_08}.

In our experiments each halo particle is treated as a test particle
which is integrated in a frozen background potential, and hence the
orbits do not self-consistently influence the frozen
potential. Nonetheless, the fact that we use a large ($1-2\times10^4$)
ensemble that is a random sampling of the halo DF gives us additional
collective power.  A DF that is not in self-consistent equilibrium
with the background potential will relax. This relaxation manifests as
orbital diffusion or mixing \citep{valluri_etal_07}.  This mixing will
occur whether the potential is time independent or varying, so long as the
DF is not in self-consistent equilibrium with the potential. We hypothesize that the diffusion rates
$\log(\Delta f)$ can be used to measure the rate of diffusion of
mixing of orbits in ensemble that is evolved in a selected
potential. If the DF is out of equilibrium we should obtain larger
diffusion rates. We test this hypothesis by comparing the distributions
of the diffusion rates of ensembles of orbits evolved in different
potentials.

Our objective is to distinguish quantitatively (with an assigned
statistical confidence) between the correct potential and incorrect
potentials using halo stars for which all 6 phase space coordinates
are available. We computed diffusion rates for all $10^4$ orbits
selected from 3 models SA1, LA1 and IA1, when the  orbits were
integrated in the correct potential. We also integrated the orbits
from the DF of SA1 in two ``slightly incorrect'' potentials: SA1
rotated about the $z$ axis by 10$^\circ$ (model SA1-10$^\circ$) and by
90$^\circ$ (model SA1-90$^\circ$).  Two other ``slightly incorrect''
potentials consisted of $10^4$ orbits from drawn from the DF of models
LA1 and IA1 and integrated in the potential for SA1 (models LA1-in-SA1
and IA1-in-SA1 respectively).

We also considered one ``strongly incorrect''  model: the orbits of the
initial triaxial halo A, were integrated in a potential consisting of
halo A + a short-axis disk. Since the halo was not allowed to relax in
response to the presence of the disk, the orbits are strongly out of
equilibrium since,in effect, the disk potential is ``turned on suddenly''.

\begin{figure}
\centering
\includegraphics[trim=2.pt 0.pt 0.pt 0pt,width=0.4\textwidth]
{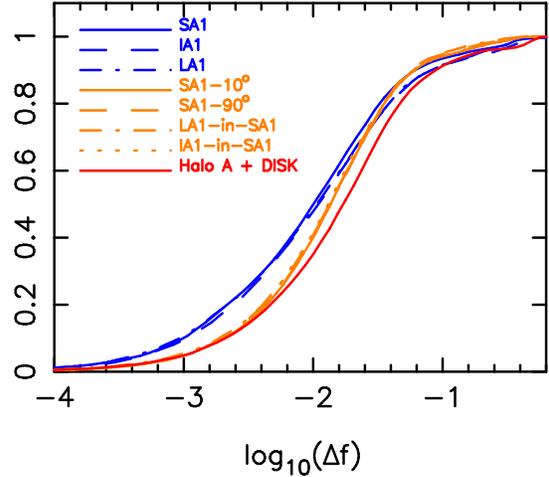}
\caption{Cumulative distribution functions of orbital diffusion rates
  $\log(\Delta f)$ for ensembles of $10^4$ orbits which were evolved
  in the correct potential (blue curves), evolved in slightly
  incorrect potentials (orange curves) and in a strongly incorrect
  potential (red).}
\label{fig:diffusion}
\end{figure}

Figure~\ref{fig:diffusion} shows the cumulative distribution functions
(CDFs) of the diffusion parameters for ensembles $10^4$ orbits from
the 3 DFs from SA1, LA1 and IA1 evolved in their own potentials as
blue curves. The CDFs of diffusion rates for $10^4$ orbits in each of
the four ``slightly incorrect'' potentials, SA1-10$^\circ$,
SA1-90$^\circ$, LA1-in-SA1 and IA1-in-SA1 (orange curves), and the
``significantly incorrect'' potential (red curve). The clear
separation of the curves shows that the CDFs of $\log(\Delta f)$ for
ensembles integrated in the correct potential (blue curves) are always
to the left of the orange curves implying that in the correct
potentials there are significantly more orbits with low diffusion
rates ($\log(\Delta f) \lesssim -2$) than when these ensembles are evolved in incorrect potentials
(orange and red curves).
 
We carried out pair-wise comparisons of the 8 CDFs in
Figure~\ref{fig:diffusion} using the non-parametric
Kolomogorov-Smirnov (KS) test to evaluate the probability $p$ that the
distributions are statistically different from each other. We compute
the KS-statistic which measures the ``distance'' $d$ between the two
distributions. Small distances $d$ between pairs of distributions and
low (or zero) values of $p$, indicate a low probability that the two
distributions are (statistically) identical. Results of representative
subsets of the tests are reported in Table~\ref{tab:KStest}. The first
three tests in the table compare the CDF of orbits from the SA1 DF
evolved in the correct potential with the same orbits evolved in three
incorrect potentials. The small values of $d$ and $p$ indicate a very
low probability that the blue curve for SA1 and the orange (or red
curves) are identical.  The next two tests compare the DFs of LA1 and
IA1 integrated in the correct potential (blue curves) with the CDFs
when the same ensembles are evolved in the potential for SA1 (orange
curves).  These two tests also yielded small values of $p$ implying a
low probability that the CDFs of LA1 and IA1 were identical to those
of LA1-in-SA1 and IA1-in-SA1 respectively. Thus the first 5 KS tests
shows that the probability of CDFs of $\log(\Delta f)$ arising from
the correct and incorrect potentials being identical is very low ($p
\sim 0$).

The next 3 tests compared various ``slightly-incorrect distributions''
(orange curves in Figure~\ref{fig:diffusion}) with each other. The
results of the KS-tests in Table~\ref{tab:KStest} show that it is
difficult to statistically discriminate between the various slightly
incorrect potentials because the probabilities of their being
identical are quite large $p > 0.1$

The last two tests compared two ``slightly-incorrect'' models (orange
curves) with the ``strongly-incorrect'' models (red curve) (labeled
``Halo A + DISK''): the small values of $p$ and $d$ indicate that 
the red curve can be statistically distinguished from the orange
curves with high confidence.
     
\begin{table}
\begin{centering}
\begin{tabular}{lll}\hline 
\multicolumn{1}{l}{Models compared in KS-Test} &
\multicolumn{1}{c}{Distance $d$} &
\multicolumn{1}{c}{Probability $p$} \\ 
& & \\
\hline
SA1 vs. SA1-10 &    0.11 &	 0.\\
SA1 vs. LA1-in-SA1 &    0.10 &		 $4.2 \times 10^{-44}$\\
SA1 vs. Halo A + DISK &    0.15 &	 0.\\
LA1 vs. LA1-in-SA1 & $9.9\times 10^{-2}$& $8.6 \times 10^{-43}$\\
IA1 vs. IA1-in-SA1 & $8.6\times 10^{-2}$ & $2.7 \times 10^{-26}$\\
 &  & \\
SA1-90 vs. SA1-10 & $1.2\times 10^{-2}$&	 0.45\\
SA1-10 vs. LA1-in-SA1 & $1.7 \times 10^{-2}$& 0.12 \\
LA1-in-SA1 vs. IA1-in-SA1 & $1.9 \times 10^{-2}$ &	 0.12\\
&  &\\
SA1-90 vs. Halo A + DISK &  $8.8 \times 10^{-2}$	&$1.7 \times 10^{-34}$\\
LA1-in-SA1 vs. Halo A + DISK & $8.1 \times 10^{-2}$	&  $1.9 \times 10^{-29}$ \\
\hline
 \end{tabular} 
\caption{Results of Kolomogorov-Smirnov tests comparing cumulative distribution functions of 
$\log(\Delta f)$ from Figure~\ref{fig:diffusion} pairwise.  }
\label{tab:KStest} 
\end{centering} 
\end{table}

The tests above shows that although each particle is treated as a
``test-particle'' in the fixed background potential we are able to
harness the collective behavior of an ensemble {\it drawn from a
  self-consistent distribution function assumed to be in steady state} to statistically identify
cases where the orbit ensemble is not in equilibrium with the
background potential. Although the diffusion rate of an ensemble (in a fixed background potential) can be quantified even without self-consistent calculations, it is more correct to think of this diffusion in terms of the collective relaxation that occurs when an N-body system is out of equilibrium, rather than chaos, since these orbits are drawn from a distribution function.

This is a fundamental result that relies on the Jeans Theorem: when
the initial positions and velocities (phase space coordinates) of
orbits are not drawn from a self-consistent DF in a steady-state
equilibrium potential, most orbits are not launched on regular tori
and hence they will diffuse in phase space \citep[travel through the
  ``Arnold web'', c.f. ][]{lichtenberg_lieberman_92}. The results
above show that although there is a spread of diffusion rates in any
given ensemble. Since the CDF of the distribution for the entire
ensemble is shifted to smaller values of the diffusion rate when the
potential is close to equilibrium and becomes statistically larger
when the potential is incorrect.

From Figure~\ref{fig:diffusion} we see that the CDF of diffusion rates of a ``slightly incorrect'' model  can be clearly distinguished from the correct model. This suggests a novel  way to utilize the six phase space coordinates of MW halo stars to distinguish between various possible models for
the halo potential. However, the overlap of orange curves for various ``slightly incorrect''
models indicates that it may not be easy to distinguish between the various ``slightly incorrect'' models in a way that allows us to  progress iteratively towards a model closest to the true potential. 

 Nonetheless this method has the value that it will work best
in the region of the MW halo that is more well mixed (i.e. the inner
halo), whereas methods such as modeling the tidal tails of dwarf
satellites relies on their being less well mixed and will primarily be
applicable in the outer halo.  Note that it is not necessary to assume that all the orbits in the correct equilibrium model are regular. In fact, the CDF of
diffusion rates in Figure~\ref{fig:diffusion} shows that even in the
correct/equilibrium models there are a few orbits which have fairly
high diffusion rates.  

It is important to note that this method will not work if the net
potential is spherical (which the MW is not because $\sim$10\% of the
total mass is in the highly flattened Galactic disk), or if the
potential is assumed to be of St\"ackel form. This is because all
orbits in spherical and St\"ackel potentials are integrable by
definition (i.e. all orbits are perfectly regular), hence the
diffusion of orbits would only measure numerical errors of the
method. However, since St\"ackel potentials are idealized potentials
of primarily theoretical interest, there is no reason to expect that
the Galaxy should have such a form.  In a future paper we will refine 
these ideas to quantitatively assess the possibility of measuring the 
shape of the Milky Way's inner halo with ensembles of stellar orbits. 


\section{Discussion and Conclusions}
\label{sec:discuss}

Understanding the structure, dynamics and formation history of the MW
is a major thrust of current astronomical research. The fundamental
science driver of the burgeoning Galactic all-sky survey industry e.g.
SDSS-SEGUE~\citep{SEGUE}, APOGEE~\citep{apogee}, RAVE~\citep{RAVE},
LSST~\citep{LSST}, PanSTARRS~\citep{panstarrs},
LAMOST~\citep{LAMOST_05}, Skymapper~\citep{skymapper_07},
HERMES~\citep{hermes_AAO} and eventually {\it
  Gaia}~\citep{perryman_etal_01}) is the premise that the Galaxy and
the Local Group were formed in a manner that typifies the formation of
galaxies in the $\Lambda$CDM paradigm. Therefore understanding the
structure, dynamics, chemistry and thereby formation history of our
own Galaxy will result in stronger constraints on the complex physics
of galaxy formation.

The use of orbital frequencies of halo stars has been recognized as an
important way to identify the relics of satellites which were tidally
disrupted in the MW halo \citep{mcmillan_binney_08,
gomez_helmi_10}. In a recent paper \citet{gomez_etal_10} also showed
that when the time dependent growth of the Galaxy and observational
errors associated with {\it Gaia} are taken into account, it will be
possible to uniquely identify about 30\% of the accretion events that
occurred in the last 10~Gyrs, while the remainder of the accretion
events will likely be difficult to disentangle from a more smoothly
mixed component.  \citet{helmi_etal_11} have argued, from a comparison
of the degree of spatial variation in the distribution of stars in the
stellar halo observed in the SDSS-II survey \citep{bell_etal_08} and
the similar distributions of halo stars in cosmological $N$-body
simulations \citep{cooper_etal_10}, that there may exist a smooth
underlying stellar halo component that is much more well mixed than
material solely accreted from satellites.  The method outlined here
complements the work that focuses on disentangling the relics of tidal
streams, in that it can be applied to both well mixed and unmixed
orbits and can therefore be applied to a much larger sample of halo
stellar orbits.

The SDSS-SEGUE survey has already obtained phase-space coordinates for
over 17,000 stars within 4~kpc of the sun.  In a paper in preparation,
\citet{derris_etal_11} apply the techniques described in this paper to
the SDSS-Segue Calibration sample \citep{carollo_etal_10}, and
correlate the orbital properties of halo stars with their
metallicities and spatial distribution. They use revised distances to
the stars in the Calibration sample \citep{beers_etal_11} that
overcome recently reported distance measurement errors
\citep{schoenrich_etal_11}. Upcoming surveys will increase both the
volume of space observed and the accuracy with which distances,
proper-motions, radial velocities and metallicities are measured for
MW halo stars, significantly impacting our understanding of the
structure and dynamics of the Galaxy.  If the stellar halo does
consist of distinct components --- a well-mixed inner halo which
formed largely {\it in situ} (i.e. in potentials deeper than typical
dwarf satellite potentials) and an outer halo that was formed
primarily from the accretion of tidally stripped dwarf satellites ---
one might also expect to observe distinct differences in the orbital
populations of these two components.

One of the primary goals of current and future surveys of the MW is to
determine the shape and radial density profile of the dark matter halo
and to construct the self-consistent phase space distribution function
of the major stellar components - the thin and thick disks, the bulge
and the stellar halo \citep{binney_mcmillan_11}. A popular and
flexible method for constructing the DF of a potential is the orbit
superposition method \citep{schwarzschild_79, vandermarel_etal_98,
  cretton_etal_99, valluri_etal_04, thomas_etal_04}, which relies on
orbit libraries that uniformly sample orbital initial condition
space. In the DF of realistic triaxial potentials like the halos of
galaxies, a large fraction of boxlike orbits are resonant (i.e. have
commensurable frequencies) \citep{miralda_escude_schwarzschild_89,
  merritt_valluri_99}.  Determining the fraction of orbits associated
with resonances is therefore important for constructing the
distribution function. However, since these orbits are ``resonantly
trapped'', they densely populate very small regions of phase space,
and are likely to be under-represented in the orbit libraries used to
construct DFs (which generally sample some initial condition space
uniformly). Resonant orbits are also not adequately represented in the
DFs constructed via orbital torus construction methods, since they
are, by definition, not present in perfectly regular St\"ackel
potentials, whose orbital tori are adiabatically deformed to construct
the DF of realistic potentials \citep[e.g.][]{binney_10,
  binney_mcmillan_11}. The methods described in this paper address
this important issue in a uniquely powerful way. Rather than
attempting to guess the initial distribution of orbits required to
populate the orbit libraries, the frequency analysis of the orbits of
halo stars in a set of trial potentials can be used to construct the
frequency map, which we have shown to yield a robust picture of the global
DF, even when orbits are selected within a limited volume in the
galaxy. The measurement of the diffusion rates of the orbits then
gives an estimate of the best trial potential, and can be used as an
additional constraint in the Schwarzschild or torus construction
method. The applicability of these methods to the construction of the
full phase space distribution function of the stellar halo, will be
discussed in greater detail in future work. 

 In the majority of simulations presented in this paper we do not
  consider the possibility that the inner and outer halos could have
  angular momenta of different directions or signs.  Warps seen in
  disk galaxies, are now believed to be evidence that the angular
  momentum vectors of outer dark matter halo is misaligned from the
  disk and the associated inner halo \citep[e.g.][]{ost_bin_89,
    jia_bin_99, deb_sel_99,shen_sellwood_06,roskar_etal_10}. These
  authors argue that cosmic infall causes the angular momentum
  direction of the outer dark matter halo to be different from that of
  the inner dark matter halo.  The build up of dark matter halos via
  cosmic infall occurs due to the accretion of satellites, and such
  hierarchical infall is also believed to build up the stellar
  halo. In a companion paper \citep{valluri_etal_11_mugs} we study the
  angular momentum distributions of both star particles and dark
  matter particles in high resolution cosmological simulations of disk
  galaxies from the MUGS collaboration
  \citep{stinson_etal_10}. Preliminary results indicate that the
  angular momentum distributions of halo stars at different radial
  distances from the galactic center can be used to trace the angular
  momentum distribution of the dark matter halos at these radii. Thus
  orbital analysis of halo stars could potentially be used to measure
  the angular momentum of the dark matter halo and test the prediction
  that differential angular momentum of the halo can produce warps.
  Model TA1 (Figures~\ref{fig:shape_contours}~(right) \&
  \ref{fig:TA1_map}) show the effect of a disk inclined to the
  equatorial plane of a triaxial halo.
  Figure~\ref{fig:shape_contours}~(right) shows that the density
  contours of the inner halo become aligned with the disk, while the
  outer halo remains unchanged. In this simulation the disk was held
  rigid, but it is likely that a live disk would form a warp due to
  the misalignment of the inner and outer halo angular
  momenta. Nonetheless this simulation is illustrative of the
  possibility of using frequency maps of halo stars to detect such
  tilts in the principle axes and angular momentum directions of
  different parts of the halo since it illustrates that the orbital
  frequencies only depend on the assumed potential. 

 To conclude we summarize the main results of this paper.
\begin{enumerate}
\item{Frequency analysis of the orbits of a large representative
  sample of halo particles can be used to construct frequency maps
  which provide a compact representation of the distribution function
  of the stellar or dark matter halo. The map also give an estimate of
  the fractions of orbits in different major orbit families and
  enables easy identification of dominant resonances. The
  identification of global resonances is important for constructing
  global DFs for the MW since global resonances strongly constrain the
  DF.  The ability of frequency maps to reveal globally important
  orbit families and resonances, even with orbits selected from a
  limited volume around the sun, suggests that this method will
  provide significant input to the construction of DFs for the Milky
  Way galaxy \citep{binney_may_86}.}

\item{The adiabatic growth of a disk in a halo results in significant
  resonant trapping of halo particles with resonant orbits appearing
  clustered along narrow resonance lines on a frequency map. In a
  Cartesian frequency map of orbits in a moderately triaxial halo we
  see several resonant box orbit families. In axisymmetric halos, the
  use of cylindrical coordinates reveals a large number of resonances
  primarily between the radial frequency $\Omega_R$ and the vertical
  frequency $\Omega_z$. The strong resonant trapping seen in all cases where the disk grows
  quiescently in a pre-existing halo imply that the identification of resonances in the Milky Way's stellar halo could provide evidence for the adiabatic growth of the disk following the formation of the halo.}

\item{Resonances are found in all of the controlled simulations in
  which disks were grown inside halos, regardless of the shape of the
  halo and regardless of how the disk was oriented relative to the
  large scale orientation of the halo. It has been previously
  demonstrated that the growth of a bayonic disk inside a triaxial
  halo deforms the inner regions to make them more oblate, but the
  halos remain modestly triaxial at large radii. We find that although
  the inner regions of the halo are similar regardless of large scale
  halo orientation, the frequency maps of inner halo particles reflect
  the differences in the global orientation of the halo relative to
  the disk.}

\item{We see that halo resonances are formed by both static disks and
  live disks (those that form spiral features and bars). However, the
  coherent time dependent perturbations from a bar can result in
  scattering of the most tightly bound particles, resulting in broader
  (less well defined) resonances.}

\item{Controlled hydrodynamic simulations, in which hot gas
  distributed throughout a dark matter halo, is allowed to cool and
  form a gas disk which then forms stars, give rise to halo frequency
  maps with resonant structure, much like the adiabatic
  simulations. This shows that resonant trapping of halo stars by the
  disk are not purely a feature of idealized collisionless simulations
  in which a rigid disk is grown in place.}

\item{We find that the cumulative distribution of orbital diffusion
  rates are lower by a statistically significant amount, when a large
  ensemble of orbits is integrated in a potential in which it is in
  self-consistent equilibrium, and that the orbital diffusion rates
  are significantly larger when the potential is incorrect. This can potentially
  provide a novel way to constrain the potential of the MW directly
  from the 6-phase-space coordinates of a large sample of Milky Way
  halo stars.}

\end{enumerate}


\section*{Acknowledgments}
MV is supported by NSF grant AST-0908346. We thank Greg Stinson and
Jeremy Bailin for detailed comments on an earlier version of this
manuscript. MV thanks Tim Beers, Eric Bell and Mario Mateo for
enlightening discussions on SEGUE and other resolved-star all sky
survey data.

\bibliography{Master}
\bsp

\label{lastpage}

\end{document}